\documentclass[preprint,journal]{vgtc}            % preprint (journal style)

\onlineid{1474}

\ieeedoi{https://doi.org/10.1109/TVCG.2024.3456398}

\vgtccategory{Research}

\vgtcpapertype{application/design study}

\title{Towards Dataset-scale and Feature-oriented Evaluation of Text Summarization in Large Language Model Prompts}

\author{%
  \authororcid{Sam Yu-Te Lee}{0009-0000-2629-3954},
  Aryaman Bahukhandi,
  Dongyu Liu and 
  Kwan-Liu Ma
}

\authorfooter{
  \item Sam Yu-Te Lee, Aryaman Bahukhandi, Dongyu Liu and Kwan-Liu Ma are with University of California, Davis.
  Email: {ytlee, abahukhandi, dyuliu, klma@ucdavis.edu}
}

\abstract{
Recent advancements in Large Language Models (LLMs) and Prompt Engineering have made chatbot customization more accessible, significantly reducing barriers to tasks that previously required programming skills. 
However, prompt evaluation, especially at the dataset scale, remains complex due to the need to assess prompts across thousands of test instances within a dataset. 
Our study, based on a comprehensive literature review and pilot study, summarized five critical challenges in prompt evaluation.
In response, we introduce a feature-oriented workflow for systematic prompt evaluation.
In the context of text summarization, our workflow advocates evaluation with summary characteristics (feature metrics) such as complexity, formality, or naturalness, instead of using traditional quality metrics like ROUGE. 
This design choice enables a more user-friendly evaluation of prompts, as it guides users in sorting through the ambiguity inherent in natural language.
To support this workflow, we introduce Awesum, a visual analytics system 
that facilitates identifying optimal prompt refinements for text summarization through interactive visualizations, featuring a novel Prompt Comparator design that employs a BubbleSet-inspired design enhanced by dimensionality reduction techniques.
We evaluate the effectiveness and general applicability of the system with practitioners from various domains 
and found that 
(1) our design helps overcome the learning curve for non-technical people to conduct a systematic evaluation of summarization prompts, and (2) 
our feature-oriented workflow has the potential to generalize to other NLG and image-generation tasks. For future works, we advocate moving towards feature-oriented evaluation of LLM prompts and discuss unsolved challenges in terms of human-agent interaction.
}
\graphicspath{{figs/}{figures/}{pictures/}{images/}{./}} % where to search for the images
\DeclareUnicodeCharacter{2217}{*}
\usepackage[svgnames,table]{xcolor}

\usepackage{tabu}                      % only used for the table example
\usepackage{booktabs}                  % only used for the table example
\usepackage{lipsum}                    % used to generate placeholder text
\usepackage{mwe}                       % used to generate placeholder figures
\usepackage{mathptmx}                  % use matching math font
\usepackage{enumitem}
\usepackage{multirow}
\usepackage{multicol}
\usepackage{amsmath} 
\usepackage{amssymb}
\usepackage{algorithm}
\usepackage{float}
\usepackage{placeins}
\usepackage{algpseudocode}
\usepackage{bbm}
\algnewcommand{\LeftComment}[1]{\Statex \(\triangleright\) #1}
\setlist{nosep}
\usepackage[toc,page]{appendix}
\usepackage{stfloats}
\usepackage[skip=1pt]{caption}
\usepackage[utf8]{inputenc}

\usepackage{colortbl}

\usepackage{array}
\usepackage{tabularray}
\usepackage{soul}
\definecolor{NavyBlue}{RGB}{0,0,128}
\definecolor{dypink}{RGB}{236,0,140}

\usepackage[most]{tcolorbox}
\definecolor{cback}{RGB}{237,239,241}
\definecolor{cframe}{RGB}{185,196,202}
\definecolor{cgrey}{RGB}{102,102,102}
\definecolor{cT}{RGB}{186,230,253}
\definecolor{cF}{RGB}{229,217,242}
\definecolor{cR}{RGB}{216,247,147}
\newtcbox{\inlineTbox}[1][]{enhanced,
 box align=base,
 nobeforeafter,
 colback=cT,
 colframe=cframe,
 size=small,
 fontupper=\footnotesize,
 left=1pt,
 right=1pt,
 boxsep=0.5pt,
 #1}
 
 \newtcbox{\inlineRbox}[1][]{enhanced,
 box align=base,
 nobeforeafter,
 colback=cR,
 colframe=cframe,
 size=small,
 fontupper=\footnotesize,
 left=1pt,
 right=1pt,
 boxsep=0.5pt,
 #1}
  \newtcbox{\inlineFbox}[1][]{enhanced,
 box align=base,
 nobeforeafter,
 colback=cF,
 colframe=cframe,
 size=small,
 fontupper=\footnotesize,
 left=1pt,
 right=1pt,
 boxsep=0.5pt,
 #1}
 
\newtcbox{\nnboxx}[1][]{enhanced,
 box align=base,
 nobeforeafter,
 colback=cgrey,
 colframe=cgrey,
 colupper=white,
 size=small,
 left=1pt,
 right=1pt,
 boxsep=0.2mm,
 arc = 0mm,
 outer arc=0mm,
 #1}

 \definecolor{sampink}{RGB}{0,0,0}
 \newcommand{\sam}[1]{{\color{sampink} #1}}
 
\usepackage{marginnote} % For better margin notes

\renewcommand*{\marginnote}[1]{}

 \definecolor{aryamanpink}{RGB}{0,0,0}

\usepackage{tabularx}

\newcommand{\mc}[2]{\multicolumn{#1}{|c|}{#2}}
\definecolor{Gray}{gray}{0.85}
\definecolor{LightCyan}{RGB}{214,239,255}

\newcolumntype{a}{>{\columncolor{white}}c}
\newcolumntype{b}{>{\columncolor{Gray}}c}
 
\newcommand{\system}{\textit{Awesum}}

\makeatletter
\renewcommand\subsubsection{\@startsection{subsubsection}{3}{\z@}%
  {-1.8ex\@plus -1ex \@minus -.2ex}%
  {0.8ex \@plus .2ex}%
  {\reset@font\bfseries\sffamily\normalsize\vgtc@sectionfont}}
\makeatother

\begin{document}
\maketitle
%%%%%%%%%%%%%%%%%%%%%%%%%%%%%%%%%%%%%%%%%%%%%%%%%%%%%%%%%%%%%%%%
%%%%%%%%%%%%%%%%%%%%%% START OF THE PAPER %%%%%%%%%%%%%%%%%%%%%%
%%%%%%%%%%%%%%%%%%%%%%%%%%%%%%%%%%%%%%%%%%%%%%%%%%%%%%%%%%%%%%%%

% the only exception to this rule is the \firstsection command
\section{Introduction}
Prompting is a new way of biasing Large Language Models (LLMs) towards the desired output with natural language instructions~\cite{brown2020language}.
Recently, OpenAI's GPT Store 
opened the door for non-technical people to customize chatbots with prompts.
The low technical barriers and high customizability of prompting democratize a wide range of tasks that previously required programming skills~\cite{jiang2022promptmaker}, e.g., personalized reading and writing assistants~\cite{petridis2023anglekindling} or visualization creation~\cite{huang2024graphimind, wang2023dataformulator, ye2024genai}. 
On the other hand, composing a desired prompt (i.e., prompt engineering) is non-trivial.
Design studies in prompt engineering~\cite{zamfirescu2023johnny, jiang2022promptmaker, kim2023evallm} have pointed out that \textit{evaluation} in prompt engineering remains arduous because of five challenges in current evaluation practices, i.e., evaluation is \textit{Opportunistic, Manual, Multi-criteria, Dynamic, and Unactionable}. 

These challenges exist even when the evaluation is done on only a few test instances, and are amplified as the evaluation scales up. 
For example, a prompt designed for text summarization on a news article dataset with one thousand articles requires an evaluation on a larger test set than just a few instances to ensure its robustness. 
We refer to this kind of prompt evaluation as ``dataset scale'', which is under-explored yet challenging.
In contrast to machine learning evaluation where quantitative metrics like F1 score are typically used, the quality of a summarization prompt output is hard to capture with quantitative metrics. Traditional metrics such as ROUGE~\cite{lin2004rouge} have been criticized in many ways~\cite{deutsch2022re}, especially their inability to differentiate between state-of-the-art models, thus not suitable for evaluating summaries generated by LLM prompts.
\marginnote{$\triangle$\_1\_1}
As a consequence, people choose the test set \textit{opportunisticly}, i.e., whatever they see first or the data points that seem easy to evaluate, instead of choosing rigorous representatives of the whole dataset. Moreover, the test set is evaluated by \textit{manually} scanning through the outputs.
This introduces \textit{multiple and dynamic criteria} in prompt evaluation, where people change evaluation criteria as they see fit after the scanning.
Most importantly, this evaluation approach does not necessarily lead to \textit{actionable insights}, i.e., what kind of instruction is needed, or how should the criteria be expressed to generate the desired output. Designers would still need to go through a highly unpredictable trial-and-error prompt refinement process.

In this work, we target text summarization prompt refinement and attempt to explore the tasks that are involved in a systematic evaluation for dataset-scale prompts and the visualization designs that foster actionable insights.
We target individuals who seek to tailor chatbots for their specific needs or work requirements using prompts and aspire to efficiently perform systematic evaluations of their prompts.
Through the lens of text summarization, we seek to generalize our findings to broader real-world tasks that require dataset-scale evaluation. 
While text summarization may not be representative of all possible prompting tasks that LLMs could perform, it presents challenges that are not straightforward to solve and studying them could provide insights towards more generalizable solutions.
First, summarization is extensively studied yet NLP researchers can not agree on a robust \textbf{quality metric} (i.e., metrics like ROUGE~\cite{lin2004rouge} that seek to quantify the quality of a summary) for evaluating state-of-the-art systems. 
\marginnote{$\triangle$\_1\_2}
This indicates that quality metrics might have reached their limits in differentiating nuanced quality differences.
Second, a significant cognitive load is required to manually evaluate summaries. Designers need to read the lengthy text and the summarized text to decide the prompt performance and refinement directions, making it infeasible at the dataset scale.

Through a pilot study, we found that \textbf{feature metrics}, i.e., computational metrics that \textbf{characterize} the summary from different facets, are more desirable than quality metrics for both technical and non-technical prompt designers. 
\marginnote{$\triangle$\_1\_3}
For example, formality and complexity might be two critical characteristics (features) of a summary if the summarization goal is to produce academic-level writing. The key distinction of feature metrics is that each metric evaluates one characteristic of the generated summary, and the desired summary does not necessarily have a higher score, e.g., a complexity score suitable for kids (lower scores) might be more desirable than for professionals (higher scores). Compared to quality metrics, feature metrics provide a multi-faceted understanding of the generated summaries and a more concrete refinement direction.

Based on this finding, we 
introduce a feature-oriented workflow that involves four tasks: \textit{Feature Selection, Example Sourcing, Prompt Refinement, and Evaluation for Refinement.}
We develop \system, a VA system that implements the workflow for text summarization, incorporating computational linguistics metrics as features and intelligent agents to support the tasks.
It uses cluster visualization to provide an overview of the dataset and support the identification of ideal examples. 
It provides prompt suggestions based on established prompting methodologies to support prompt refinement.
Finally, a scatter plot inspired by BubbleSets and enhanced with dimensional reduction techniques supports users generate actionable insights for prompt refinement.
We recruit experts from various domains for practitioner review, from which we confirm the effectiveness of the system. 
We report the generalizability of the system to a broader range of tasks and discuss insights into human-agent interaction. 
Our contributions are as follows:

\begin{itemize}
    \item 
    
\marginnote{$\triangle$\_1\_4}
    We introduce a feature-oriented workflow 
    to address challenges in supporting dataset-scale prompt evaluation, 
    which we summarized from a literature review and a pilot study.

    \item We develop a VA system, \system, that supports the feature-oriented workflow for text summarization with computational linguistics, intelligent agents, and interactive visualizations. 
    
    \item We evaluate the effectiveness and generalizability of the system with a case study and interviews with practitioners from various domains, and report implications for future directions in prompt evaluation and human-agent interaction.
\end{itemize}

\begin{figure*}[t]
    \centering
    \includegraphics[width=\textwidth]{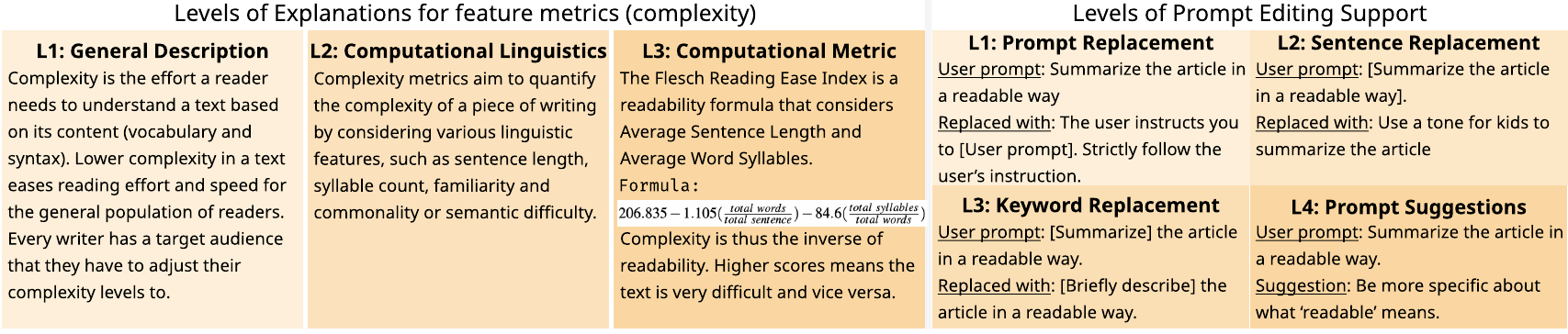}
    \caption{\textbf{Left}: Levels of explanations for feature metrics (using complexity for demonstration). L1: a textual description that vaguely defines the complexity of a text. L2: a textual description but in the context of computational linguistics, outlining common linguistic features that are considered. L3: A specific computation metric for complexity, including a textual description and the computation formula. \textbf{Right}: Levels of prompt editing support. L1: the system inserts user prompts into a template prompt under the hood. L2: the system suggests a better sentence. L3: the system suggests a better keyword or noun phrase. L4: the system gives suggestions on the prompt based on established prompting methodology. }
    \label{fig: levels}
    \vspace*{-0.5cm}
\end{figure*}
\vspace*{-0.2cm}
\section{Related Works}
In this section, we first explore the technical challenges and solutions of text summarization evaluation. We then examine current interactive visual interfaces and design studies for iterative prompt engineering, highlighting the research gap in evaluating dataset-scale prompts.

\vspace*{-0.1cm}
\subsection{Text Summarization Evaluation}
Automatic summarization is a classic natural language generation (NLG) task that converts a lengthy source text into a condensed text containing the most important information. However, the evaluation of a summarization system has remained a persistent challenge.
In the past, people predominantly used n-gram overlap metrics like BLEU~\cite{papineni2002bleu} or ROGUE~\cite{lin2004rouge} to assess the quality of a generated summary. Later, they are enhanced with embeddings-based similarity measurements like BERTScore~\cite{Zhang2020BERTScore} and MoverScore~\cite{zhao-etal-2019-moverscore}. 
Still, studies have shown that they cannot reliably quantify improvements if the difference is too small, because they are not sensitive enough to capture the subtle differences in high-quality summaries that humans can perceive~\cite{deutsch2022re}.
This became a serious problem as the capability of NLG models advanced, and finally, these metrics became ineffective as LLMs demonstrated the capability of generating human-level summaries. 

Another problem with these metrics is that they all require a labeled dataset (i.e., references) to work with, essentially transforming the summary quality evaluation into a similarity evaluation, where a human-written summary is assumed to be optimal. This is not practical in prompting as the practitioners can be non-technical and the outcome of prompting can easily exceed in quality that of any human-labeled dataset. 
Researchers have been calling for new metrics to assess the NLG quality of most recent models~\cite{novikova-etal-2017-need, bhandari-etal-2020-evaluating}. 
While some reference-free metrics that do not require labels are proposed, such as QA-QG metrics~\cite{fabbri2022qafacteval, scialom-etal-2021-questeval} or LLM-as-evaluators~\cite{zheng2023llmjudge, wang2023pandalm}, they do not yet show a consistent correlation with human judgments~\cite{shen2023large}, might be capturing spurious correlations~\cite{durmus-etal-2022-spurious}, or might exhibit various biases~\cite{tjuatja2023llms}.
% Generalizing beyond summarization, previous studies have shown that human judgment is multi-criteria and dynamic~\cite{kim2023evallm}. 
We argue that metrics that attempt to capture the overall ``quality'' of a prompt output (i.e., quality metrics) are susceptible to misaligning with human judgment, as previous studies have shown that human judgment is multi-criteria and dynamic~\cite{kim2023evallm}. 
As an alternative, we introduce \textit{feature metrics} that characterize the outputs to support sensemaking on the prompt performances.
For summarization, we use feature metrics such as formality and naturalness to support dataset-scale prompt evaluation.

\vspace*{-0.1cm}

\subsection{Interactive Visualizations for Model Refinement}
\marginnote{$\triangle$\_2\_1}
Facilitating model refinement with interactive visualizations is an extensive field~\cite{hohman2018visual}.
Many works help model developers debug their models~\cite{wongsuphasawat2017visualizing, pezzotti2018deepeyes, strobelt2018lstmvis, liu2018analyzetraining} by visualizing the data flows or training patterns, but they are designed specifically for the underlying model architecture and are not applicable in prompt engineering.
Works that focus on model tuning~\cite{golovin2017google, mljar, sigopt, wang2019atmseer} are more similar to the settings in prompt engineering, where visualizations such as line charts or parallel coordinates~\cite{golovin2017google} are used to visualize the performance metrics and their relations with hyperparameter settings. 

Still, new visualization techniques are needed in prompt engineering for three reasons.
First, previous systems are designed for model developers. In prompt engineering, the target users are extended to non-technical people whose knowledge on machine learning can not be assumed. 
Second, previous systems evaluate performances with quantitative metrics such as the F1 score, but these quantitative metrics are not applicable in most prompt evaluation scenarios.
Third, previous systems guide users by visualizing the relations between certain hyperparameter settings and the performance metrics.
In prompt engineering, the search space is all possible text expressions that can not be enumerated, and the non-deterministic nature of prompting introduces a high uncertainty in the outputs, making it hard to identify relations between prompts and performances.
In our work, we explore the possibility of evaluating prompt performances with feature metrics, which are more approachable for non-technical users and can guide them in discovering the relations between prompts and performances.

\vspace*{-0.1cm}
\subsection{Interactive Prompt Engineering and Design Studies}
\marginnote{$\triangle$\_2\_2}
Given its uniqueness, several design studies have been conducted to explore the challenges in prompt engineering~\cite{jiang2022promptmaker, zamfirescu2023johnny} and evaluation~\cite{kim2023evallm}.
Zamfirescu et al.~\cite{zamfirescu2023johnny} found that prompt evaluation practices of  non-technical users are \textit{opportunistic} rather than systematic, due to the lack of experience in controlling automatic systems.
Jiang et al.~\cite{jiang2022promptmaker} found the mental load of \textit{manually} skimming large volumes of text introduces a significant evaluation challenge. Moreover, it is hard to transform the evaluation into prompt refinement, i.e., \textit{little actionable insight is generated}.
Kim et al.~\cite{kim2023evallm} focused on prompt evaluation and reported two additional challenges. First, evaluation is \textit{multi-criteria}, i.e., the quality of the outputs could not be evaluated with a single criterion. 
Second, evaluation is \textit{dynamic}, i.e., designers expand or change their criteria  as they observe unexpected flaws in the outputs. 

Many works have attempted to support prompt engineering in an interactive, code-free environment. 
PromptIDE~\cite{strobelt2022promptide} and PromptIterator~\cite{suvcik2023prompterator} support the iterative experimentation process. As researchers and practitioners introduce more prompting techniques and best practices~\cite{promptengineeringuide, openaiprompt}, Kim et al.~\cite{kim2023cells} and Arawjo et al.~\cite{arawjo2023chainforge} propose to design chains of reusable blocks to separate prompt design, model selection, and evaluation. 
Another line of work focuses on providing prompt suggestions, ranging from word-level~\cite{mishra2023promptaid}, sentence-level~\cite{brade2023promptify}, to prompt-level~\cite{macneil2023promptmiddleware}, where users prompts are replaced with an expert prompt template. 

We summarize the prompt evaluation challenges as \textit{Opportunistic, Manual, Multi-criteria, Dynamic}, and \textit{Unactionable}.
Previous design studies and applications either focus on evaluation at the instance scale, i.e., the prompt is refined and tested on at most a few instances, or support only evaluation at the dataset scale in supervised settings, where performances can be measured with loss or accuracy with labeled datasets.
In our work, we explore the tasks and designs to support text summarization evaluation at the dataset scale, which is under-explored and calls for new visualization techniques.

\vspace*{-0.2cm}
\section{Pilot Study}\label{sec: pilot_study}
From the literature review, we have learned that quality metrics are ineffective in text summarization evaluation~\cite{deutsch2022re, novikova-etal-2017-need, bhandari-etal-2020-evaluating} and hypothesized that feature metrics, such as complexity and naturalness, could be a new way of dataset-scale prompt evaluation. 
As prompting is a relatively new research area and the typical workflows and challenges are not well-studied, 
we conducted semi-structured interviews with 6 participants to verify findings from existing works~\cite{kim2023evallm, jiang2022promptmaker, zamfirescu2023johnny}, explore new challenges that emerge at dataset scale, and confirm the feasibility of using feature metrics for dataset-scale prompt evaluation.
We recruited users of LLMs from both technical and non-technical backgrounds (e.g., ChatGPT). 
Below, we report our study design and findings.

\vspace*{-0.1cm}
\subsection{Study design}
In our interview, seek to answer the following research questions (\textbf{RQs}):
\begin{itemize}[leftmargin=*]
    \item \textit{\textbf{RQ1}: What are the challenges in dataset-scale summarization prompt evaluation?}
    \item \textit{\textbf{RQ2}: Can feature metrics address the challenges in \textbf{RQ1}?}
    \item \textit{\textbf{RQ3}: What is the preferred level of explanation of feature metrics?} 
    \item \textit{\textbf{RQ4}: What is the preferred way of supporting prompt optimization?}
\end{itemize}
To provide context for the participants, we used summarization on a news article dataset as a simulated scenario in the interview. For \textbf{RQ1}, participants were first asked to write an initial prompt that can summarize a news article, and then brainstorm ways to evaluate it. Then, we introduced a set of potential features to be used for evaluation, and then asked participants to brainstorm ways of evaluation again (\textbf{RQ2}). To answer \textbf{RQ3} and \textbf{RQ4}, we prepared different levels of feature metric explanations and prompt editing support, as shown in~\autoref{fig: levels}, and asked the participants to choose levels that they prefer and explain why. 

\vspace*{-0.1cm}
\paragraph{Participants}
We recruited participants with varying backgrounds: 
two data/visualization scientists (\textbf{P1-2}), two NLP researchers (\textbf{P3-4}), and two non-technical researchers with backgrounds in environmental science (\textbf{P5-6}).
They also have varying levels of experience with prompts. The most experienced participant is an LLM researcher who has worked on multiple related research projects and an internship. Middle-level experienced participants (N=2) had designed prompts programmatically. The least experienced participants (N=3) had used ChatGPT for preparation of presentations, or summarizing long texts.
Participants reported frequent usage of prompts for at least half a year. 

\vspace*{-0.1cm}
\paragraph{Procedure}
The 30-minute interview was conducted in a semi-structured way. We started by asking about the participants' background, experience with prompting, and their typical workflow. Then, we introduced a simulated scenario in which they designed a summarization prompt for news articles, that takes \textbf{one} news article and outputs \textbf{one} summary and emphasized that the prompt should be generalizable to 100 articles.
Participants were not required to write actual prompts since we were only interested in their thought process. Finally, we asked about their preferences on different levels of feature metrics and prompting support. All participants received a 5 USD compensation. 

\begin{figure*}[t]
    \centering
    \includegraphics[width=\textwidth]{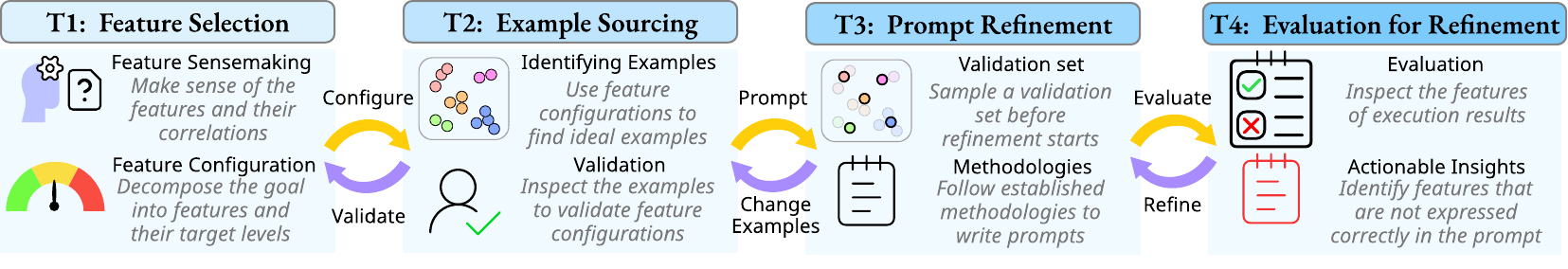}
    \caption{The feature-oriented workflow for a systematic dataset-scale prompt evaluation. \inlineTbox{T1}: Feature Selection. Designers should make sense of potential features to use for evaluation and their correlations, and then select a feature configuration that consists of a subset of features and corresponding target ranges. \inlineTbox{T2}: Example Sourcing. Designers should use the feature configuration to find ideal examples, which are used to validate (and refine) the configuration and in the prompt. \inlineTbox{T3}: Prompt Refinement. First, select a validation set before the refinement starts, and then follow established prompting methodologies to write the prompt that expresses the feature configurations. Execute the prompt for validation when ready. \inlineTbox{T4}: Evaluation for Refinement. Inspect the evaluation result on selected features, identify cases that do not fit the feature configuration, and refine the prompt accordingly. We emphasize that features should be used as guidance, not targets. }
    \label{fig: workflow}
    \vspace*{-0.5cm}
\end{figure*}

\vspace*{-0.1cm}
\subsection{Findings}
Regardless of their backgrounds, all participants reported a similar workflow for prompting: first write an initial prompt to get a baseline response, and iteratively refine the prompt by skimming through the responses. 
% \rev{The reported challenges agree with the five challenges summarized in previous works. Below, we report findings that are not covered by existing literature.}h
The reported challenges align with the five challenges we summarized from the literature review. In addition, we identified several new findings when prompts are evaluated at the dataset scale:

\noindent\inlineFbox{F1} \textbf{Dataset-scale evaluation is infeasible without support.}
The simulated scenario revealed that some external support is necessary for dataset-scale evaluation (\textbf{RQ1}). 
In their current practice, participants would randomly pick a few examples to see if the prompt works well.
% Such instance-level evaluation could be done opportunistically and manually, however demanding it might be.
% even though it is a demanding process. 
All participants (\textbf{P1-6}) agree that such a way of evaluation becomes infeasible as the evaluation scales up to a dataset as they would need to manually skim through hundreds of summaries, far beyond the physical cognitive limit of human beings. Some (\textbf{P3, P4}) suggested automated metrics as a possible solution, but \textit{``it is hard to find a metric that can generalize across the diverse tasks enabled by prompting''}.
The non-deterministic nature of prompting makes it hard to guarantee that the prompts can generalize well without actually executing the prompts, which would be costly in time and slow down the iterative refinement. 

\noindent\inlineFbox{F2} \textbf{Features are critical to dataset-scale evaluation.}\quad
Decomposing goals into features before the refinement presents opportunities to address the dynamic and unactionable challenge.
When asked to evaluate prompts that generate an ``academic'' summary, participants were clueless about the suitable evaluation criteria. However, after introducing the feature metrics, all participants agreed that these metrics could be used as evaluation criteria to cover various summarization requirements (\textbf{RQ2}), e.g., ``academic'' could be expressed as complex, very formal, and having a neutral sentiment.
Previous works have pointed out that prompt evaluation is \textit{dynamic} in that designers frequently change the evaluation criteria and redefine ``success'' after finding unexpected flaws in the output, and that it is \textit{hard to gain actionable insights}~\cite{kim2023evallm}. 
Our interview revealed that not knowing which features constitute the intended goal is a major obstacle in evaluation.
By assisting designers to systematically make sense of potential features and select the most appropriate ones, the evaluation criteria are less likely to change, and designers can identify weak aspects in the iterative refinement process.

\noindent\inlineFbox{F3} \textbf{Metrics are guidance, not target.}\quad
When asked about the preferred explanation of feature metrics (\textbf{RQ3}), participants predominantly (\textbf{P2-6)} chose L2: \textit{textual definitions in computational linguistics} (\autoref{fig: levels}). 
Most participants excluded L1 for being too generic and L3 for being too specific for the model to follow as an instruction.
Moreover, some (\textbf{P2, 4, 6}) had doubts about the reliability of the L3 computational formula, questioning that \textit{``it might not be a good representation of complexity''}. \textbf{P2} emphasizes that \textit{``(designers) should not optimize for the formula, because LLMs can understand complexity more deeply''}.
This observation echoes with GoodHart's Law~\cite{strathern1997improving}: \textit{when a measure becomes a target, it ceases to be a good measure}, and is reflective of the current situation in text summarization, where researchers are discouraged from using metrics like ROGUE to evaluate LLM-generated summaries.
As a result, we assign each feature with semantically meaningful categorizations and refrain from showing metric values. 

\noindent\inlineFbox{F4} \textbf{Prompting methodology is more important.}\quad
All participants preferred prompt suggestions over prompt replacements (\textbf{RQ4}),
despite the different levels of prompt editing support introduced in previous works.
\textbf{P6} commented that \textit{``(receiving suggestions) is a more learnable experience''}. Other participants also expressed the desire to \textit{learn} to design good prompts, i.e., the methodology of prompting~\cite{promptengineeringuide, openaiprompt}. 
We also found that even without knowledge of the established methodologies, participants developed similar methodologies through their own prompting experience, such as chain of thoughts (CoT)~\cite{wei2022chain} or providing examples.
This suggests that the design of prompting support should communicate the underlying prompting methodologies.

\vspace*{-0.2cm}
\section{Requirement Analysis}
Combining the literature review and the pilot study findings, 
we identified four tasks that users must perform in a systematic evaluation workflow: Feature Selection, Example sourcing, Prompt Refinement, and Evaluation for Refinement, as shown in~\autoref{fig: workflow}. 
Since the workflow is guided by features, we refer to it as a \textit{feature-oriented} workflow.
Next, we introduce each task and discuss its necessity in detail.
\subsection{Systematic Workflow}
\begin{enumerate}[label=\inlineTbox{T\arabic*}]
    \item \textbf{Feature Selection:} Designers need to make sense of the potential features and their correlations, and then pick a subset that best represents the intended goal. Designers should avoid fixating on minor feature values \inlineFbox{F3}, and instead set target ranges (configurations) of the features.
    As indicated by \inlineFbox{F2}, making Feature Selection explicit and systematic enables the subsequent evaluation to become less dynamic and more predictable. 
    \item \textbf{Example Sourcing:}  Once the features are selected, designers should find ideal examples that fit the features, which can be used to validate the feature configurations. Examples are also used in the prompts or compare the outputs in subsequent tasks. 
    \item \textbf{Prompt Refinement:} To guarantee a systematic evaluation, designers need to also refine prompts systematically. First, a representative \textit{validation set} must be chosen statistically to guarantee its validity. Then, prompting should follow the methodologies proposed by prompt engineers and researchers to write better and more controllable prompts \inlineFbox{F4}. Finally, prompt drafts are executed on the validation set to generate insights for refinement.
    \item \textbf{Evaluation for Refinement:} With previous tasks, we establish a base to ensure that a systematic evaluation is possible.
    After getting feedback from the selected features,
    we emphasize that designers should use features as guidance, not targets \inlineFbox{F3}. Once the refinement on the validation set is completed, designers can test the prompt on the whole dataset to confirm the performance.
\end{enumerate}
Ideally, designers would complete the tasks sequentially (\autoref{fig: workflow}, yellow arrows).
Nevertheless, the workflow supports an iterative process: designers can use the feedback from a later task to refine a previous task (purple arrows). For example, if \inlineTbox{T3} shows that the examples are not representative, designers can go back to \inlineTbox{T2} for better examples. 
Such a systematic workflow presents a structured way of refining and evaluating prompts, allowing easy identification of failure points.

\vspace*{-0.15cm}
\subsection{Design Requirements}\label{sec: design_requirements}
A systematic evaluation with the above tasks is non-trivial for prompt designers to do properly \inlineFbox{F1}, especially for non-technical people. 
We summarized four design requirements to support \inlineTbox{T1--4}:
\begin{enumerate}[label=\inlineRbox{R\arabic*}]
    \item \textbf{Support sensemaking and recommendation of feature metrics}. As our target audience does not necessarily come from a technical background, selecting features introduces a steep learning curve. Our system should support an automatic recommendation of feature metrics based on the user's goal and provide explanations. 
    \item \textbf{Support overview and identification of ideal examples.} 
    Identifying and validating ideal examples at the dataset scale is cognitively demanding. To facilitate this process, our system should use statistical analysis to guarantee the soundness of the examples and interactive visualizations to reduce the cognitive load.
    \item \textbf{Provide suggestions based on established methodologies.} Prompting is a relatively new area and methodologies are constantly improving. As our users might not be aware of such advancements, our system should be designed to enforce state-of-the-art methodologies and give suggestions when necessary. 
    \item \textbf{Support visual tracking of prompt refinement effects.} It is critical yet demanding for designers to track the evaluation results with pure statistics. Our system should support such tracking through visualization and convey the performance of each prompt. 
\end{enumerate}

\section{\system: System Design}
\begin{figure*}[t]
    \centering
    \includegraphics[width=\textwidth]{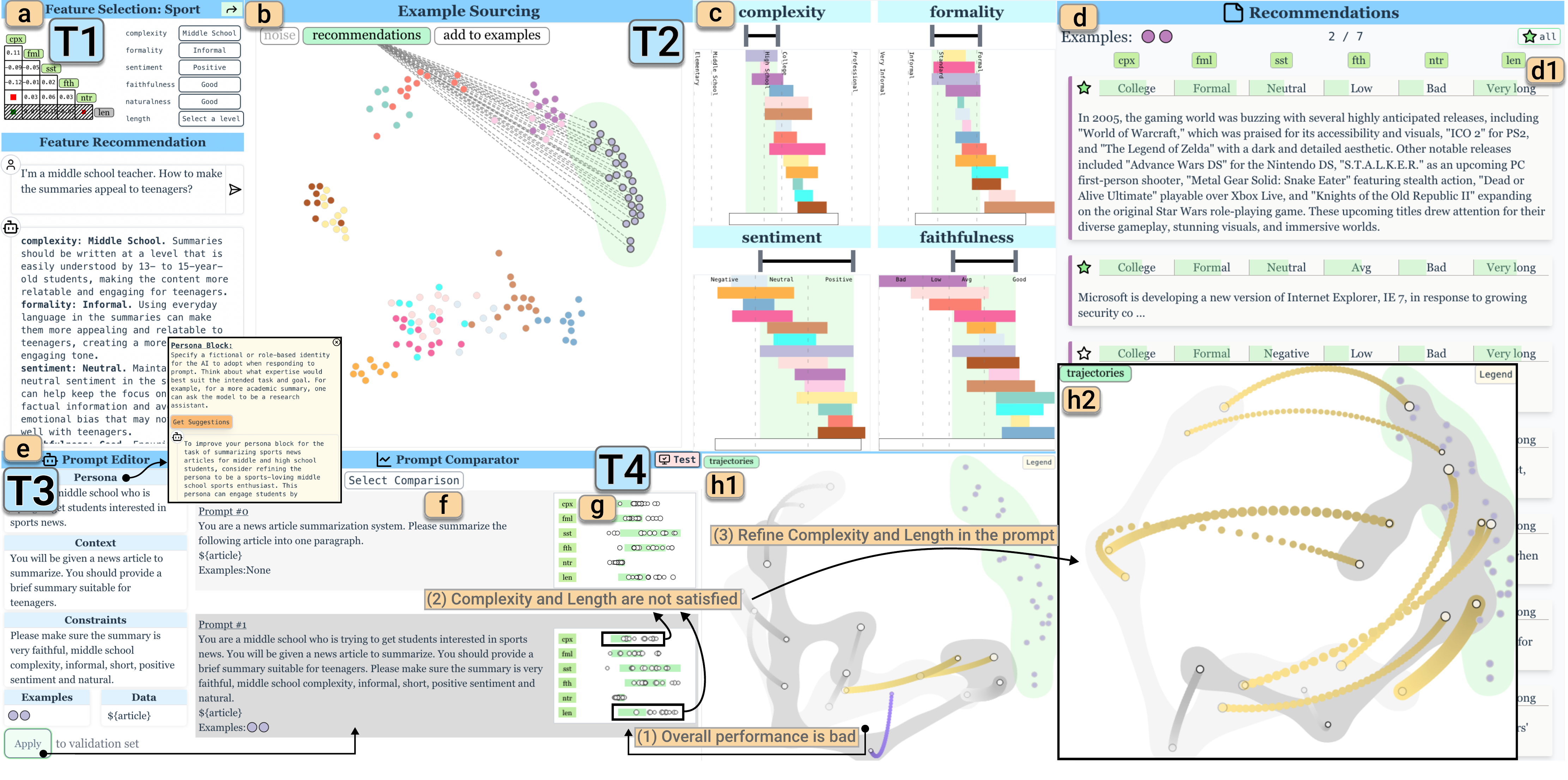}
    \captionsetup{aboveskip=0pt}    
    \caption{Actions done by Alice in the case study. 
    Alice uses Feature Selection View (a) to decompose the summarization goal ``appeal to teenagers'' into feature configurations \inlineTbox{T1}.
    After setting the configuration, the system finds the closest cluster in (b) and highlights it in a green bubble. She uses Recommendation View (d) to validate the examples and adjust the feature configuration in (c) \inlineTbox{T2}. 
    Then, she moves on to Prompt Editor View (e) and writes the first prompt according to the feature configuration \inlineTbox{T3}, with the help of the prompt suggestions given by a chatbot.
    She executes the first prompt and to compare it with the baseline prompt in Prompt Comparator View (f), which provides visual tracking of all the versions of prompts. 
    From Bubble Plot (h1), she can see that the prompt is performing badly, as the dark and light gray bubbles are overlapping and not close to the green bubble.  
    She goes to the dot plot (g) and finds that complexity and length are not satisfied, as many validation cases are falling out of the green bars. 
    She refines the prompt by adjusting how she expresses complexity and length \inlineTbox{T4} by taking prompting suggestions from the system, and the new prompt yields much better performance (h2), as the dark gray bubble overlaps significantly with the green bubble, and most of the curves are yellow.
    }
    \label{fig: case_study}
    \vspace*{-0.4cm}    
\end{figure*}

Based on the design requirements, we developed \system\footnote{\url{https://github.com/SamLee-dedeboy/Awesum}}, a visual analytics system that supports the feature-oriented workflow on dataset-scale summarization prompt refinement and evaluation (\autoref{fig: case_study}). 
Next, we introduce the details of each component in detail.

\vspace*{-0.2cm}
\subsection{Feature Computation}
\system \ characterizes summaries with six features and guides users toward their goals. We use well-established metrics for complexity, formality, and sentiment, but no known metrics suit our needs for faithfulness and naturalness, so we introduce new computations for them.  Following \inlineRbox{R1}, we categorize each metric into easy-to-understand levels, encouraging users to not fixate on minor differences in metric values. Below, we briefly summarize their definitions. Computational details and categorizations are presented in the appendix.

\vspace*{-0.15cm}
\paragraph{\textit{\textbf{\footnotesize Complexity}}} The complexity score characterizes the ease with which a reader can understand a written text. We use the Flesch Reading Ease Index~\cite{kincaid1975derivation} to measure the complexity of a text based on sentence count, word count, and syllable count. 
We simplify its original eight-level categorization into five levels: \textit{Elementary, Middle School, High School, College}, and \textit{Professional}, which indicates the knowledge level required for a reader to easily understand the text.

\vspace*{-0.15cm}
\paragraph{\textit{\textbf{\footnotesize Formality}}} The formality score characterizes how formal a piece of text is in terms of linguistic structures, conventions, and vocabulary. We use the measure of textual lexical diversity (MTLD)~\cite{mccarthy2010mtld}, which is calculated from the ratio of unique word stems to the number of words. We categorize it into \textit{Informal, Standard, Formal} and \textit{Very Formal}.

\vspace*{-0.15cm}
\paragraph{\textit{\textbf{\footnotesize Sentiment}}} The sentiment score characterizes the emotional tone expressed in a piece of text. We use VADER~\cite{hutto2014vader}, a lexical-based sentiment analysis model to measure the sentiment of a text, which generates a score between $[-1, 1]$. Then we categorize the sentiment into \textit{Negative, Neutral,} and \textit{Positive} using a threshold of 0.3. 

\vspace*{-0.15cm}
\paragraph{\textit{\textbf{\footnotesize Faithfulness}}} Faithfulness characterizes the degree to which a generated text is consistent with the input information in terms of semantic similarity, completeness, and accuracy~\cite{liu2024cliqueparcel}.
We incorporate this feature in response to the ``hallucination'' issue that exists in most LLMs and is of the most concern to prompt designers. 
Although more advanced approaches exist, we calculate the faithfulness score based on Named Entity Recognition overlap (NER-overlap)~\cite{laban2022summac} for its transparency, robustness, fast computation speed, and reasonable alignment with human judgment.
We categorize it into \textit{Bad, Low, Avg, } and \textit{Good}. 
% Specifically, we conduct fuzzy-matching between entities in the original content and the summary and compute the ratio between the matched entities and the original ones as the faithfulness score. 

\vspace*{-0.15cm}
\paragraph{\textit{\textbf{\footnotesize Naturalness}}} Naturalness characterizes how well a text reads like human-written. To the best of our knowledge, no metrics have been proposed to evaluate the naturalness of texts. Based on the insights from Pu et al.~\cite {munoz2023contrasting} that LLM-generated text exhibits statistical differences in certain linguistic features, such as part-of-speech (POS) tags, 
we conducted experiments on a dataset~\cite{zhang2301benchmarking} that contains both human-written and LLM-generated summaries to select differentiable linguistic features and use them to compute the naturalness score with a weighted sum.
The score is similarly categorized into \textit{Bad, Low, Avg} and \textit{Good}.

\vspace*{-0.15cm}
\paragraph{\textit{\textbf{\footnotesize Length}}} We use word count as the length of a text. Although a simple feature, prompt designers and our pilot study participants have reported difficulties in controlling the length of the generated text. We thus include it and categorize it into \textit{Short, Mid, Long, } and \textit{Very Long}.

Finally, we construct a feature vector $F=(f_1, f_2, f_3, \dots)$ for each summary, where $f_i$ is the numerical score of a feature calculated on the generated summary. 
Then for each feature, we compute its z score as the value of each dimension in the feature vector. 
% All features are then standardized by removing the mean and scaling to unit variance. 
The feature vector of each summary is the basic computation unit in other parts of the system, namely correlation analysis, clustering, and dimensionality reduction.

\marginnote{$\triangle$\_5\_1}
We chose the above six feature metrics because there is a clear semantic meaning to each feature. Considering that our target users could include non-technical people, the interpretability of the metrics is essential to a systematic and rigorous evaluation. Moreover, we refrain from using LLM-evaluators to support user-defined features~\cite{kim2023evallm}, even though it might broaden the applicability of the system, for three reasons. First, LLM evaluators have received criticism for their inconsistent alignment with human judgment~\cite{shen2023large} and potential biases~\cite{tjuatja2023llms}. Second, we have conducted experiments to show that the non-deterministic nature of LLMs makes them unreliable in dataset-scale evaluation. Experimentation details are presented in supplemental materials. Third, the goal of developing \system \ in this paper is to verify the effectiveness of the feature-oriented workflow on text summarization task. Considering the above limitations of LLM evaluators, it might introduce unnecessary confounding factors in the evaluation. 

\vspace*{-0.1cm}    
\subsection{Feature Selection}
\textit{Feature Selection View} supports \inlineTbox{T1} with two sub-tasks: feature correlation analysis and feature recommendation \inlineRbox{R1}, which supports users to set feature configuration for their goals of the summarization prompt. 

\begin{figure}[t]
    \centering
    \includegraphics[width=\columnwidth]{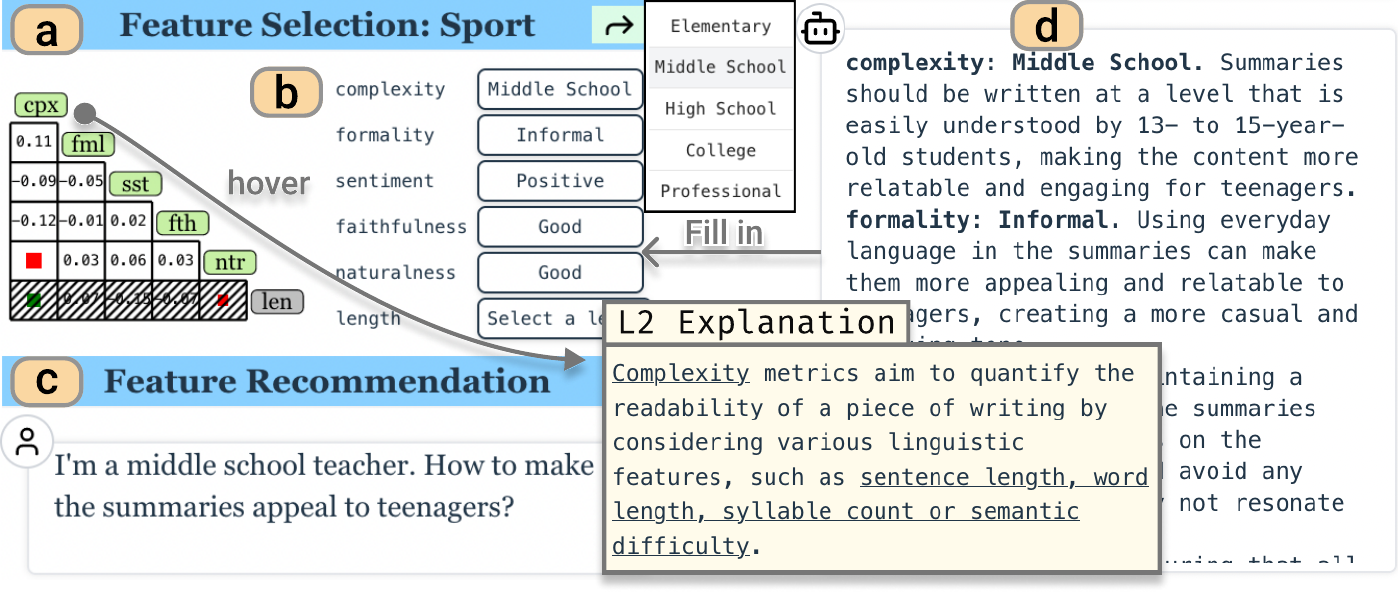}
    \caption{Feature Selection View is designed to support \inlineTbox{T1}. The feature correlation matrix (a) facilitates sensemaking of the feature definitions and their correlations. If the correlation exceeds a certain threshold, it is encoded by a red (negative) or green (positive) square. The rows and columns are sketched out in stripes if the user deems a feature irrelevant to the goal. 
    Then, users can manually set the feature configuration in (b) using the dropdown menus, or use the feature recommendation chatbot (c) to get recommendations. The chatbot responds in (d) with a recommended configuration by considering the L2 definition of each feature and the user's goal and automatically fills them in (b). }
    \label{fig: feature_selection}
    
    \vspace*{-0.5cm}
\end{figure}

\vspace*{-0.1cm}    
\paragraph{\textit{\textbf{\footnotesize Feature Correlation Matrix}}}
Feature correlation matrix shows the Pearson correlation coefficients between all pairs of features (\autoref{fig: feature_selection}-a), which is designed to prevent users from selecting conflicting features that are hard to accomplish simultaneously.
% which are critical for selecting the feature configuration, as users might select conflicting features that are hard to accomplish simultaneously. 
For example, if the correlation matrix suggests that complexity and naturalness have a negative correlation, users should not pick a configuration that has both high (or low) complexity and naturalness. 
Visualizing the correlations as a matrix allows users to avoid such conflicts at a glance \inlineRbox{R1}. 
The L2 explanations of each feature can be inspected by hovering over the feature tags.
Based on the result of the baseline prompt, the system calculates all the feature vectors $F$ on the generated summaries and the Pearson correlation coefficient between all pairs of features. If the coefficient exceeds a threshold, we use a square to indicate its significance and encode the correlation strength with the length. Negative and positive correlations are encoded by red and green, respectively. Features omitted by the user in the configuration are shaded in stripes. 
\marginnote{$\triangle$\_5\_2}

\vspace*{-0.1cm}
\paragraph{\textit{\textbf{\footnotesize Feature Recommendation}}}
 In case the feature correlation matrix does not provide a concrete idea for the user, \textit{Feature Recommendation Panel} (\autoref{fig: feature_selection}-c and -d) integrates a chatbot to recommend feature configurations. Users can express their goals in natural language (e.g., generate summaries suitable for academic writing), and the chatbot will respond with a recommended configuration with explanations \inlineRbox{R1}. Under the hood, we provide the L2 feature definitions and their categorizations to the chatbot to ensure reasonable responses. If significant correlations exist in the features, the chatbot would highlight the correlations and give suggestions accordingly.
The system automatically fills in the recommended configuration (\autoref{fig: feature_selection}-b). Dropdown menus are provided to change feature levels manually.
\vspace*{-0.1cm}

% \vspace*{-0.3cm}
\vspace*{-0.1cm}
\subsection{Example Sourcing}\label{sec: example_sourcing}
\begin{figure}[t]
    \centering
    \includegraphics[width=\columnwidth]{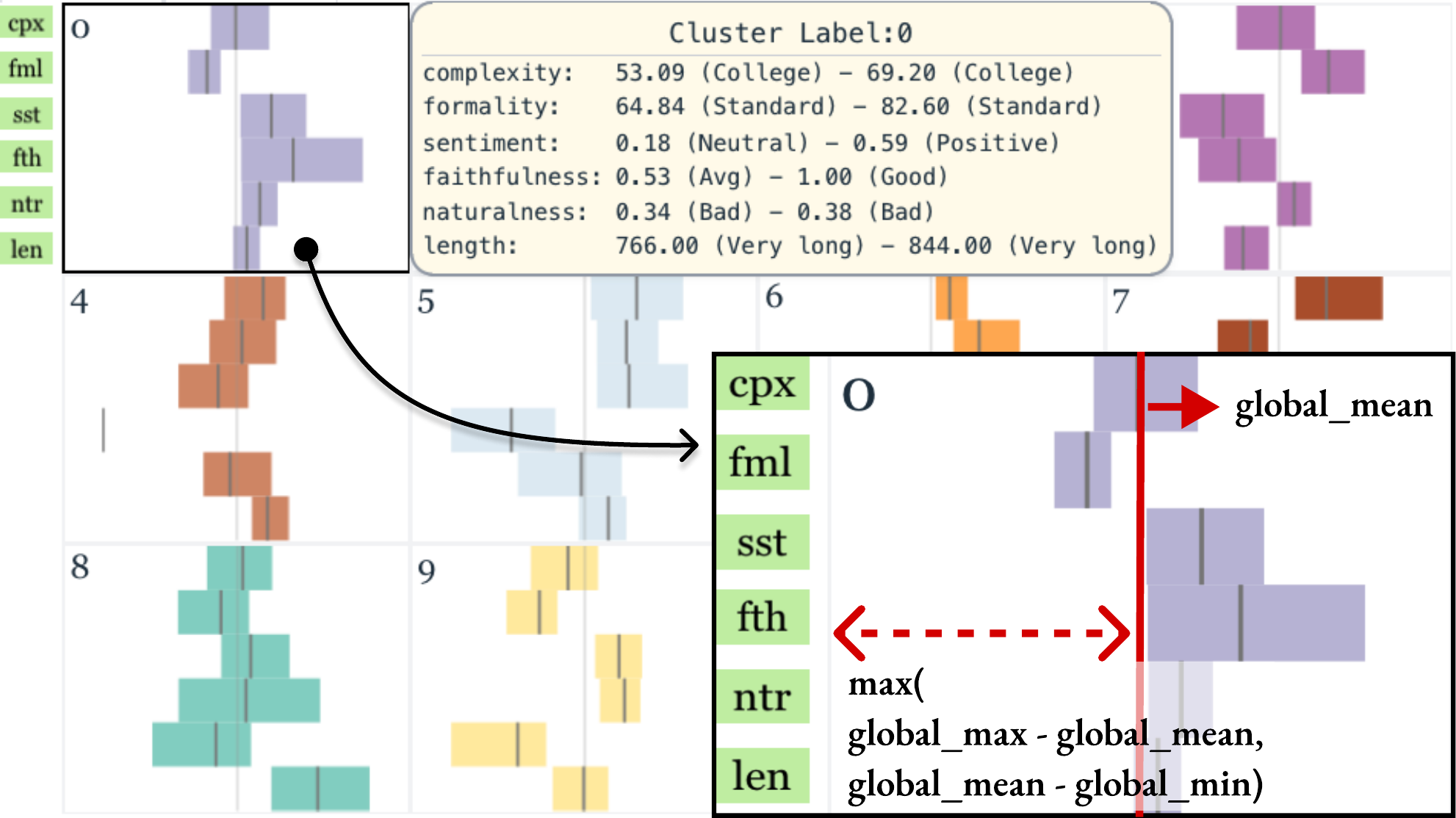}
    \caption{
       Cluster Profiles visualize the feature ranges of each cluster with a scaled vertical bar chart. Each cluster is encoded by a unique color. The bars are aligned by the center vertical line, which encodes the global mean. The half of the profile width is encoded by $max(global\_max - global\_mean, global\_mean - global\_min)$. This way, the bars are scaled to show the distinctiveness of each cluster. Hovering over a bar chart shows the ranges in values and categorization. 
    }%
    \label{fig: cluster_stats}
    \vspace*{-0.4cm}    
\end{figure}

\textit{Example Sourcing View} supports \inlineTbox{T2} with Cluster Plot (\autoref{fig: case_study}-b), which presents an overview of the dataset, and a side panel that can be switched between Cluster Profiles (\autoref{fig: cluster_stats}), which visualizes cluster feature characteristics; and Feature Distributions (\autoref{fig: case_study}-c), which provides finer control over feature ranges selection \inlineRbox{R2}.

\marginnote{$\triangle$\_5\_3}
Users can inspect Cluster Profiles or Feature Distributions and narrow down the target feature ranges (green bubble in~\autoref{fig: case_study}-b and green ranges in~\autoref{fig: case_study}-c) to identify ideal examples. 
\vspace*{-0.1cm}
\paragraph{\textit{\textbf{\footnotesize Cluster Analysis}}}
The system applies the OPTICS clustering algorithm~\cite{ankerst1999optics} on the initial summaries using their feature vectors. The OPTICS algorithm has several benefits over other clustering algorithms such as KMeans~\cite{lloyd1982kmeans}. First, OPTICS is more flexible as it automatically detects the cluster densities to decide cluster numbers and thus does not require prior knowledge of the dataset's cluster shapes. Second, it ensures that all generated clusters have low variance in feature vectors, making each cluster distinctive.
Third, it removes ``noises'' that are not close to any clusters, which is ideal for identifying examples as our users might not have the cognitive power to identify examples from a large amount of data. These benefits make the OPTICS algorithm ideal for our system.
In addition, The clustering results are used to generate the validation set by under-sampling~\cite{yen2009clustersampling}, i.e., cluster centroids are assigned to the validation set, ensuring the diversity of the validation set for robust evaluation.
\vspace*{-0.1cm}
\paragraph{\textit{\textbf{\footnotesize Cluster Plot}}}

\marginnote{$\triangle$\_5\_4}
In Cluster Plot, each initial summary is encoded as a circle and colored by their corresponding cluster.
\system \ applies Kernel PCA~\cite{scholkopf1997kernel} with cosine distance on the feature vectors of the initial summaries to generate 2D coordinates, which plots clusters with similar feature characteristics closer to each other, forming regions in 2D space that represent certain characteristics.
The design considers maintaining the visual continuity between example identification and tracking prompt refinement effects (\autoref{sec: prompt_comparator}). Thus, we look for \textit{parametric} dimensionality reduction methods, where an explicit mapping function (i.e., projection to low-dimensional space) can be reused, excluding popular non-parametric methods like t-SNE~\cite{van2008tsne}, UMAP~\cite{mcinnes2020umap} or MDS~\cite{cox2000mds}.
Kernel PCA allows us to reuse the same projection in subsequent steps where a visual tracking of the performances of different prompts is provided, while capturing the non-linear feature relationships in the data.
We additionally employ two techniques to improve visual clarity. First, we hide noise points deemed by the OPTICS algorithm as they are not representative, and only show them upon user toggling. Second, we apply collision detection and a force-directed layout that attracts each point to its cluster's centroid. Although this would affect point positions given by Kernel PCA projections, this is acceptable as the exact position of each point would not affect users identifying ideal examples \inlineTbox{T2}. Since our target audience is non-technical prompt designers, we emphasize clarity over accuracy. 

\vspace*{-0.15cm}
\paragraph{\textit{\textbf{\footnotesize Cluster Profiles}}}
Cluster Profiles show the feature ranges of each cluster in a scaled vertical bar chart (\autoref{fig: cluster_stats}). Considering \inlineRbox{R2}, we want to highlight distinguishing feature characteristics among clusters through Cluster Profiles. Inspired by the Difference Overlay design~\cite{srinivasan2018barcomparison}, we scale the bars by aligning the global mean of each metric at the center of a profile. Then, we take the maximum of $(global\_max - global\_mean, global\_mean - global\_min)$ as the range of half of the width and scale each bar accordingly. This way, users can easily identify the distinctive features of each cluster. Clicking on a cluster profile highlights its position in Cluster Plot (green bubble) and automatically sets the cluster points as ideal examples. 
\vspace*{-0.15cm}
\paragraph{\textit{\textbf{\footnotesize Feature Distributions}}}
Feature Distribution Panel (\autoref{fig: case_study}-c) supports users in identifying ideal examples with finer control over the feature configurations \inlineRbox{R2}. It shows the unscaled ranges of each cluster for each feature, which complement the scaled ranges in Cluster Profile while maintaining the same visual encoding for colors (cluster label), emphasizing visual continuity and simplicity. The cluster bars are ordered from top to bottom by their mean value on the corresponding feature. For each feature, users can click on a cluster bar to set its range as a target or drag the double-direction slider at the top to adjust the target range, indicated by a light green background. 
Supporting \inlineTbox{T2} at a finer scale ensures the system adapts to diverse goals.

\begin{figure}[t]
    \centering
    \includegraphics[width=\columnwidth]{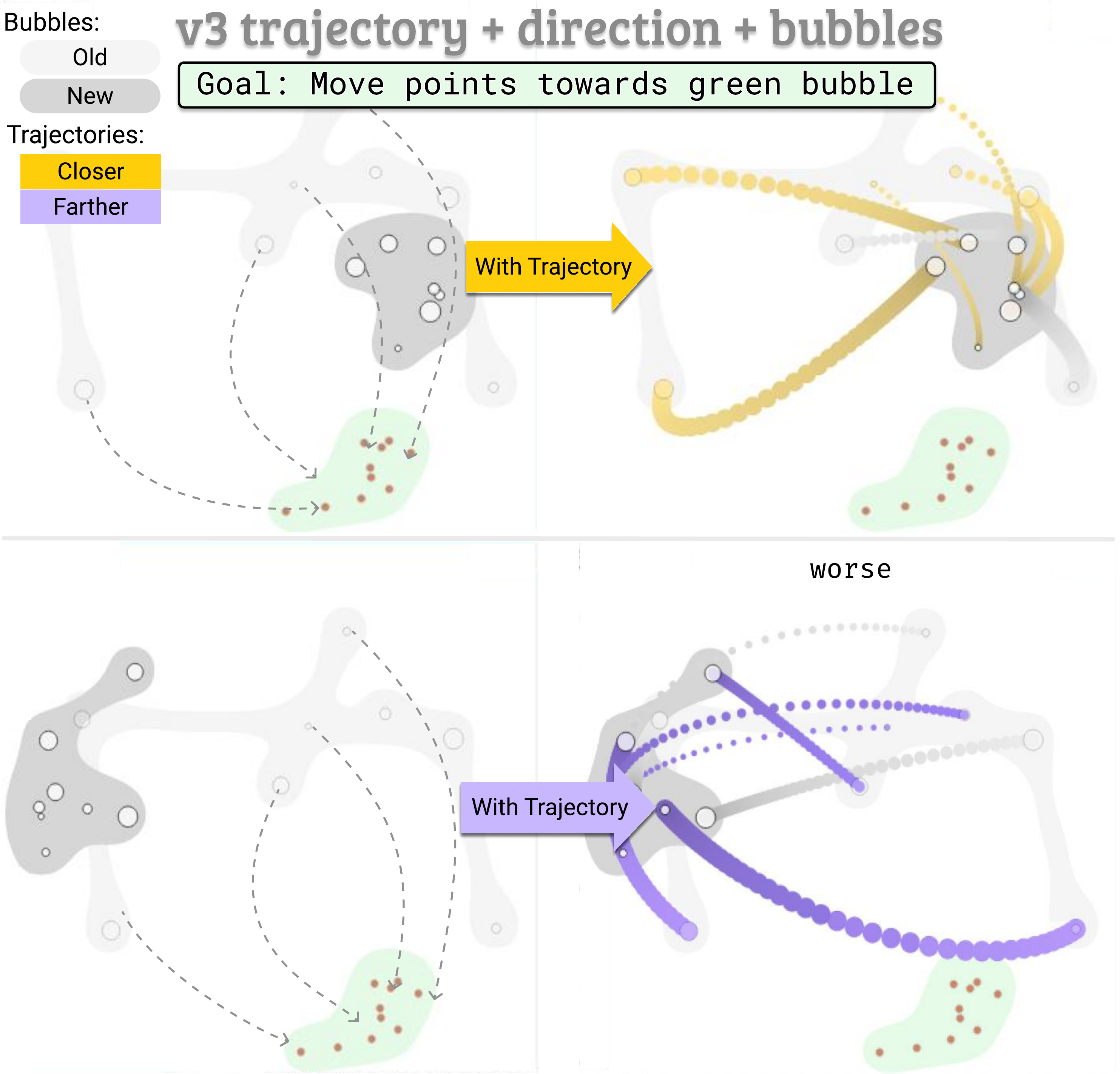}   \captionsetup{aboveskip=3pt}
    \caption{Final design of Bubble Plot. The light green bubble represents the goal. Light and dark gray bubbles (representing old and new prompts) allow users to estimate prompt performances by the distances from the green bubble. 
    A visual pattern that indicates a ``better'' prompt (top) would have the dark bubble moving closer to the light green bubble, and vice versa (bottom). 
    Colored curves connect validation cases in old and new prompts, with color encoding changes in feature values (yellow for better, purple for worse, gray for insignificant).
    The importance of a validation case is encoded by the circle size (and trajectory width). 
    }
    \label{fig: bubble_patterns}
    \vspace*{-0.5cm}
\end{figure}

\vspace*{-0.1cm}    
\subsection{Recommendation}
\textit{Recommendation View} (\autoref{fig: case_study}-d) supports validation of the target ranges of the features \inlineTbox{T2}. It shows the content of the examples with their feature categorizations. 
After inspecting the examples, users can click the star icon and add them to the prompt. 

The design of Recommendation View considers two weaknesses in Example Sourcing View. First, it is hard for users to estimate the number of recommended examples in Cluster Plot, as users would need to estimate the region's area to estimate the number. To complement, we use a fixed height for each example in Recommendation View by collapsing excessive content, ensuring as many examples are shown in the viewport as possible. This allows users to estimate the number of examples at a glance with the total height of the examples. 
Second, Cluster Profiles are scaled so they do not strictly encode the range of each feature. This could mislead the users into choosing the wrong feature configuration. To complement, we encode the feature values in horizontal green bars (\autoref{fig: case_study}-d1). Users can skim through the examples with awareness of the true ranges of each feature at a glance. 
\vspace*{-0.1cm}    
\subsection{Prompt Editor}
\textit{Prompt Editor View} supports users with Prompt Refinement \inlineTbox{T3}. Informed by state-of-the-art prompting methodologies~\cite{promptengineeringuide, openaiprompt}, we divide a prompt into five blocks: \textit{Persona, Context, Constraints, Examples} and \textit{Data}. In each block, users have a clear goal for the content of the section \inlineRbox{R3}. For example, Persona block should \textit{specify a role-based identity for the AI to adopt}. 
This explicit division of a prompt enforces the users to follow established prompting methodologies with a minimum requirement for knowledge of the methodologies. 
Once a prompt is written, users can click the ``Apply'' button to test it on the validation set. 
In addition, users can hover over the block titles to learn the purpose of the block and get suggestions specific to their feature configuration from a chatbot \inlineRbox{R3}, as shown in ~\autoref{fig: case_study}-e.

\begin{figure}[t]
    \centering
     \includegraphics[width=\columnwidth]{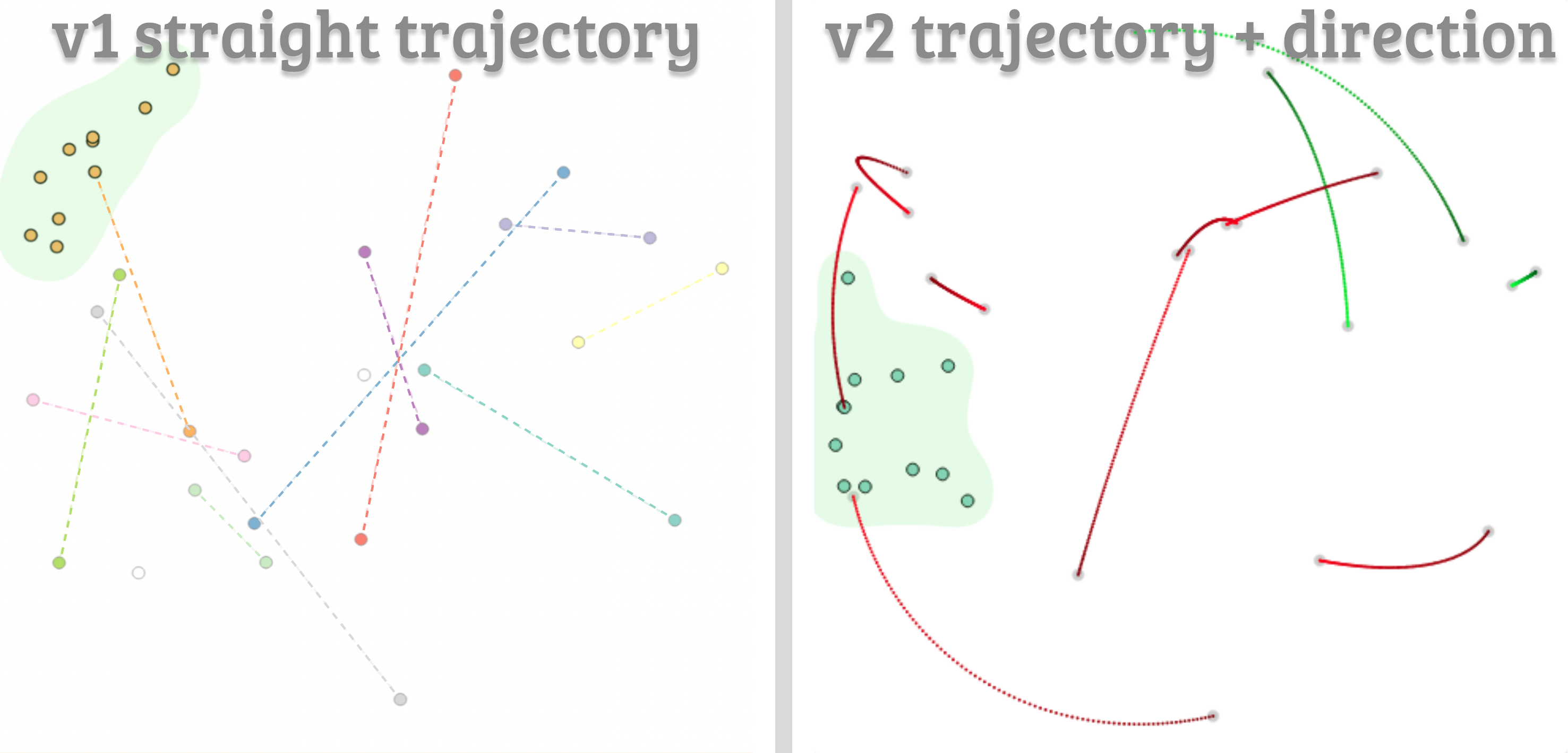}
    \captionsetup{aboveskip=4pt}    
     \caption{Previous designs of Bubble Plot. \textbf{v1}: Validation cases are encoded by cluster colors. Animated dotted straight lines connect validation cases from old and new prompts. It is cluttered and hard to gain any valuable insight. \textbf{v2}: Cluttering is reduced by using curves to connect points, which are generated by sampling in high-dimensional space. Colors are based on the directions (green for better, red for worse). The brightness (light to dark) encodes the connection direction (old to new prompt). The visual pattern highlights comparison, but it is hard to make sense of the performance of individual prompts. }
     \label{fig: preivous_bubble_plot}
    \vspace*{-0.6cm}    
\end{figure}

\vspace*{-0.1cm}
\subsection{Prompt Comparator}\label{sec: prompt_comparator}
\textit{Prompt Comparator View} supports \inlineTbox{T4} with two components: Prompt Tracking Panel (\autoref{fig: case_study}-f) and Bubble Plot (\autoref{fig: bubble_patterns}). Both components incorporate visualizations to reduce the cognitive load of keeping track of multiple iterations of prompts and their performances \inlineRbox{R4}. 
Since the refinement is conducted on the validation set,  a ``Test'' button is provided to execute the prompt on the whole dataset once users are satisfied with the prompt. Next, we introduce each component in detail.
\vspace*{-0.3cm}
\paragraph{\textit{\textbf{\footnotesize Prompt Tracking}}}
Prompt Tracking Panel (\autoref{fig: case_study}-f) stores all the prompts that the user has written, including the prompt content, the examples, and their evaluation result using a horizontal dot plot (\autoref{fig: case_study}-g). 
Tracking the history of prompts is essential as prompt refinement and evaluation is a highly iterative process. 
To provide visual tracking \inlineRbox{R4}, each prompt snippet includes a dot plot that visualizes the performance of the prompt in detail, where each row is a feature and each dot in a row is a validation case.
Since the validation set is generated from the cluster centroids, we encode the dot size with its corresponding cluster size to indicate its importance, i.e., larger dots are more important because they represent more points. 
We stroke the feature configuration in light green on the dot plot to clearly present the goal of the prompt: make all the dots fall into the light green bars.

\vspace*{-0.1cm}
\paragraph{\textit{\textbf{\footnotesize Bubble Plot}}} 
Bubble Plot (\autoref{fig: bubble_patterns}) supports visual comparison between any two prompts and visualizes their ``distances'' from the ideal examples to support \inlineTbox{T4}.
As illustrated in~\autoref{sec: example_sourcing}, we reuse the Kernel PCA projection in Cluster Plot for all points in the plot to ensure visual continuity. 
Ideal examples are highlighted in a light green bubble as an indication of the goal. 
Two iterations (old and new) of the validation cases are plotted as circles, where the circle radius encodes its corresponding cluster size to suggest importance, and are surrounded in light and dark gray bubbles, respectively.
This allows the user to estimate the performance of a prompt through the distance between its bubble and the green bubble \inlineRbox{R4}.
The expected visual pattern for a ``better'' prompt is that the dark bubble is ``moving towards'' the green bubble, and vice versa.
To make the comparison more explicit, we connect the same validation case from two versions of prompts with curves generated with linear sampling in high-dimensional space. This effectively visualizes the ``trajectory'' of the validation case incurred by the new prompt. Specifically, for each pair of feature vectors $(F_1, F_2)$, we linearly sample $\epsilon=100$ points between $F_1$ and $F_2$ and project each point to 2D space, thus forming the curve and encoding the importance with curve width.
\marginnote{$\triangle$\_5\_5}

Since the projection is non-linear, the sampled trajectory appears to be curved. 
Also, using a fixed number of sampling points ($\epsilon$) makes long trajectories appear dotted. 
We use a color-blind-friendly color palette to indicate the direction of the trajectory: yellow if the validation case is moving closer to the ideal examples, and vice versa, purple. 
Validation cases with non-significant changes (below a threshold) are colored in gray.
Users can click the ``Select Comparison'' button to select any two iterations of prompts to compare. 
\vspace*{-0.2cm}
\paragraph{\textit{\textbf{\footnotesize Bubble Plot Design Iteration}}}
The design of Bubble Plot went through three iterations, as shown in~\autoref{fig: preivous_bubble_plot}.
In \textbf{v1}, we encoded the cluster of each validation case with color and connected validation cases from two different prompts with an animated straight dotted line to indicate their moving directions. However, the lines were cluttered, and interactions were needed for further analysis. Barely any insights can be generated at a glance. Inspired by Scheepens et al.~\cite{scheepens2016animatedflow}, in \textbf{v2}, we mitigated the clutter issue by connecting validation cases with curves generated by linearly sampling in high-dimensional space and used green (closer) and red (further) to encode the moving direction from the validation cases. Users can now compare two prompts at a glance, but it was hard to estimate the performance of individual prompts, which is critical in \inlineRbox{R4}.
This motivated us to surround the validation cases with BubbleSets to indicate the relative distance from the target bubble, providing a clear visual tracking of prompt performance.

\vspace*{-0.1cm}
\section{Case Study}
We demonstrate how a non-technical prompt designer with no prior prompting experience can use \system \ to refine and evaluate prompts. 
% We highlight the low technical barrier of the system and the benefit of following the workflow.

Alice is a middle school teacher and she wants to customize a chatbot that can summarize sports news in a way that appeals to teenagers.
She loads the sports-related materials into \system \ and uses a baseline prompt to generate the initial summaries. The system executes the prompt and processes the summaries.
Alice starts by selecting the features that make a summary appeal to teenagers \inlineTbox{T1}. She asks in Feature Recommendation Panel: ``I'm a middle school teacher. How to make the summaries appeal to teenagers?''. The chatbot answers with a recommended feature configuration with explanations. She checks the definitions of the features and adjusts ``sentiment'' to ``positive'' as she wants the summaries to be more energetic, as shown in \autoref{fig: case_study}-a. 

After setting the feature configuration, she then proceeds to find ideal examples that fit this configuration \inlineTbox{T2}. She clicks the arrow in the green background at the top right, which triggers the system to automatically find a cluster that best fits the feature configuration. The system highlights the purple cluster in a green background and dotted lines that connect to ``recommendations'' at the top (\autoref{fig: case_study}-b).
She skims through the content and the feature ranges of the purple cluster in Recommendation View (\autoref{fig: case_study}-d) to ensure that they fit her expectations. She is satisfied with the cluster and selects the two best examples to use in the prompt (starred in green). 

After Feature Selection and Example Sourcing, Alice proceeds to write the prompt \inlineTbox{T3}. Even without prior prompting experience, she can easily understand what each block means and what should be written as the blocks resemble interaction patterns with chatbots. She clicks the ``Get Suggestions'' button to brainstorm some ideas for Persona block (\autoref{fig: case_study}-e), then decides to put ``You are a middle school teacher who is trying to get students interested in sports news'' in Persona block. She repeats similarly for Context block and Constraints block and then executes the prompt on the validation set.
The system executes the prompt and recalculates the features. The new prompt's content and its evaluation result are presented in Prompt Tracking Panel (\autoref{fig: case_study}-f), with the new prompt colored in dark gray and the baseline prompt colored in light gray. 
From Bubble Plot (\autoref{fig: case_study}-h1), she can quickly tell that the new prompt has a bad overall performance as the gray bubble is not moving closer to the green bubble.
She inspects the dot plot (\autoref{fig: case_study}-g) and finds that \textit{Complexity} and \textit{Length} are not being satisfied. She thus goes back to the prompt and refines the parts where she describes complexity and length \inlineTbox{T4}, seeking the prompt suggestion chatbot for help. 
After the refinement, the new prompt is much better as the dark gray bubble has significant overlaps with the green bubble with only a few exceptions. 
The trajectories show that most validation cases are moving towards the green bubble, and only a few do not have significant changes.
Alice is satisfied with this prompt and she clicks the ``Test'' button to test it on the whole dataset, which shows that the prompt works as expected.

This case study shows that the system is effective in supporting dataset-scale prompt refinement and evaluation. Using the system, even non-technical people with no prior prompting experience could write a reasonably good prompt, evaluate the prompt, and figure out which parts need to be refined. 
By following the systematic feature-oriented workflow behind the system, users overcome the first four challenges (opportunistic, manual, multi-criteria, and dynamic) in prompt evaluation. Cluster Plot and Bubble plot allow users to overcome the unactionable challenge by visualizing the prompt performance.

\vspace*{-0.2cm}    
\section{Practitioner review and Discussion}
\system \ is evaluated through a practitioner review. Even though the system is designed for text summarization, we aim to explore a broader range of tasks in people's everyday work that can also be supported. We recruit practitioners from various backgrounds who are interested in customizing LLMs to automate their professional workflow. We evaluate the effectiveness of the system with them and discuss its generalizability. In addition, we report challenges in prompt evaluation and human-agent interaction that arise in professional usage.
\vspace*{-0.2cm}
\subsection{Participants and Procedure}
Four practitioners from different backgrounds were recruited:
C1 is a correspondent for the New York Times specializing in scientific topics; C2, C3, and C4 are researchers in ecology, NLP, and sociology, respectively. 
C1, C2, and C4 came from non-technical backgrounds and had only prompted in ChatGPT.
All participants expressed strong interest in customizing LLMs with prompts to automate their workflow. 

The review procedure consists of three sessions, 20 minutes each. In the first session, we introduce the background and give a quick walkthrough of the interface. In the second session, participants choose a topic (from Politics, Sport, Technology, and Business), propose a summarization goal, and then use \system \ to write prompts that fulfill that goal. Finally, participants engage in a semi-structured interview. We provide at least 25 US dollars in compensation for all participants.

\vspace*{-0.15cm}
\subsection{User Feedback}
Overall, the system received positive feedback on usability and user-friendliness. Still, a minimum amount of training is needed to use the system. Below, we report the practitioner review on system design, prompting methodology, learning curve, and visualization literacy.
\vspace*{-0.1cm}
\paragraph{\textit{\textbf{\footnotesize System Design}}}
% Participants can quickly understand the workflow and the functionality of each component after the tutorial.
The system was highly praised for being easy to follow despite its complexity.
Participants commented that \textit{``I have no trouble understanding what I should do (C2)''} and that \textit{``The organization makes a lot of sense (C1)''}.
All participants successfully wrote a well-performed prompt for their goals based on the feedback from the system. 
As \textbf{C3} said, \textit{``You can tell (from Bubble Plot) that initially my prompt wasn't good, then I changed the prompt according to the features (L2 descriptions) and it's giving exactly what I wanted.''}
Bubble Plot and dot plots facilitated participants to examine and locate unfulfilled features and the L2 descriptions helped them refine prompts.

\vspace*{-0.1cm}
\paragraph{\textit{\textbf{\footnotesize Prompting Methodology}}}
Providing suggestions supported by the established prompting methodologies inspired less experienced participants to explore more possibilities, going beyond our initial expectation that simply facilitates prompt writing: \textit{``(C4): If I had just a blank text box, I would not know how to write the first sentence \dots and I would not know that this is something that you can do. ''}
Most inexperienced prompt designers find prompting challenging because they do not know what the model is capable of. Before using the system, C4 did not know that designers could instruct the model to adopt a persona, and that inspired her to explore what could be used as a persona. As opposed to L1--3 prompt editing support (\autoref{fig: levels}), providing prompting suggestions (L4) taught participants what to expect from the model.

\vspace*{-0.1cm}
\paragraph{\textit{\textbf{\footnotesize Learning Curve}}}
There is still a non-trivial learning curve to overcome, despite being highly praised for its clarity. For C1, \textit{``without you to talk me through (the interface), it would have been hard to figure out, (because) there is a ton of information onto one page.''}
This shows that for people from non-technical backgrounds, engaging in a systematic workflow is not something they are familiar with and could be challenging at the beginning. 
Still, the clarity of the workflow and visualization design helped smooth this process, as C1 and C4 both commented that \textit{``it was like playing a game''}.
By visualizing the prompt's performance with Bubble Plot, the system provided a clear goal and abundant visual hints for the participants to grasp their current status and how far away they were from the goal, leading to a less unpredictable experience.
\vspace*{-0.1cm}
\paragraph{\textit{\textbf{\footnotesize Visualization Literacy}}}
For people unfamiliar with high-dimensional visualization, Cluster Plot could be confusing. As C2 mentioned, \textit{``I can understand that each region has its own characteristics and that the green bubble is my goal, but I would still want to know how it works under the hood to really make sure I'm not misunderstanding.''}
C4 also raised a similar concern: \textit{``Figuring out what these functionalities do is one thing; figuring out what these functionalities mean,
that's another thing.''}
This is a valid concern since Kernal PCA introduces distortion in the two-dimensional scatter plot, which could mislead users who are not aware of it. 
Even though we designed Bubble Plot with loose encodings to mitigate this issue and encouraged users to use metrics as guidance, users could still fixate on absolute metric values without a minimum amount of training. 
% C4 is the only participant who has graduate-level training in visualization research, implying a higher visualization literacy.

\vspace*{-0.15cm}
\subsection{Design Implications}
In this section, we first discuss how the system can be generalized to a broader range of tasks. Then, we report three implications on human-agent interaction: the issue of trust, experimenting with LLM capabilities, and using features as boundary objects.
\vspace*{-0.1cm}
% \subsubsection{Generalizability}
% \paragraph{\textit{\textbf{\footnotesize Evaluating Information-condensing Prompts}}}

\paragraph{\textit{\textbf{\footnotesize Supporting Information-Condensing Tasks}}}
\marginnote{$\triangle$\_7\_1}
Beyond summarization, using the system to evaluate prompts for condensing information is a highly praised usage.
When asked about how well the system could be generalized to other tasks in their professional work, all participants agreed its strong potential in condensing information, e.g., generating logical relations for multiple articles (C1), leading to the importance of proper evaluation, as C4 mentioned: 
\textit{``Say I take 2000 tweets, something I might do with AI is to clean these tweets because not all tweets have meaningful things according to my use\dots but I can't use it unless I have a grasp of what the AI's assumption is about `meaningful'.''
}

Challenges in evaluating information condensing tasks resemble those in text summarization in many ways: there is a great amount of information (data); it is cognitively demanding to evaluate even a small amount of outputs; and it is probably unreliable to evaluate all of them with quality metrics.
Participants agreed that the current feature metrics in \system \ most scenarios for their professional work, which is reasonable because the standards for ``professional writing'' are quite similar.
In general, the system is shown to readily apply to information-condensing tasks.

\paragraph{\textit{\textbf{\footnotesize Shift in Task Formulation}}}
\marginnote{$\triangle$\_7\_2}
We observe a shift in the formulation of researchable tasks from practical tasks in the practitioner review. For example, to understand a large amount of customer feedback quickly (practical), researchers in traditional machine learning formulate the task as sentiment analysis or topic modeling task (researchable). 
In the pilot study and practitioner review, non-technical participants can design prompts that directly solve the practical task at hand. In prompt engineering, it is less important to rigorously define a researchable task given the high customizability of prompts.
Therefore, while the system seems not applicable to many existing NLP tasks, such as data extraction~\cite{li2023evaluateChatgpt}, making it less advantageous for NLP researchers, it still brings much value to non-technical practitioners.

\vspace*{-0.1cm}

\paragraph{\textit{\textbf{\footnotesize Applying to Image Generation Tasks}}}
\marginnote{$\triangle$\_7\_3}
C3 was very positive about applying the system to image generation tasks: \textit{``you would need some other metrics (for image generation) like the styles of the images, but the other things are more or less the same.''}
The system design makes it easy to extend feature metrics for a broader range of tasks.
Given the right set of features, the rest of the system is agnostic of the underlying features because OPTICS clustering and Kernel PCA both operate on feature vectors.
For image generation prompt evaluation, a potential direction for computing feature metrics is to evaluate the style of an image through image similarity analysis~\cite{palubinskas2017image}. For example, one could collect a dataset of cartoon images and calculate the similarity between a generated image and the dataset to quantify how ``cartoony'' the image is. Such stylistic metrics could be used as feature metrics to evaluate image generation prompts.
In general, the system has the potential to be extended for image generation tasks, but developing corresponding feature metrics is out of scope and we leave it for future work.

% \vspace*{-0.1cm}
Beyond generalizability, participants also brought up topics related to human-agent interaction in their professional work.

\vspace*{-0.1cm}
\paragraph{\textit{\textbf{\footnotesize Experimenting with LLM Capabilities}}}
By observing the changes in features, designers can experiment with prompts to better understand the capability of LLMs, as shown by two unexpected usages by C3 and C4 during the practitioner review. C4 tried paraphrasing, e.g., changing the persona from ``reporter'' to ``editor'', and observing changes in relevant features, such as complexity and formality.
For inexperienced prompt designers, coping with the ambiguity inherent in natural language has been one of the reasons why prompting is challenging. 
Experimenting with prompts mitigates this issue as it guides designers in sorting through the ambiguity: \textit{``It's actually not putting stuff in unambiguous form, but still less ambiguous enough that you can track where things are going. (C4)''}
C3 used the system for a more technical experiment, where he tested the positional bias~\cite{wang2023positionbias} of LLMs. C3 put conflicting instructions in different parts of the prompt, e.g., asking the model to adopt the persona of ``an elementary school teacher writing for kids'' but generate summaries ``as academic as you can'', and observing which part of the instruction is followed based on the features. 
Both usages showed that evaluation through features has the potential for understanding and even diagnosing LLMs. 

\vspace*{-0.1cm}
\paragraph{\textit{\textbf{\footnotesize Features as Boundary Objects}}}
The practitioner review revealed the potential to use features as boundary objects~\cite{star1989institutional}, i.e., a common ground that could be reached between humans and agents in prompting.
C2 and C4 both felt that using natural language to communicate with AI agents is different from communication with humans. When asked why, C4 replied that \textit{``I know what I want, but how do I say it has to do with context''}, suggesting that context needs to be presented before proper communication can continue with each new conversation. C2 pointed out that the ability to sort through vague communication is critical: \textit{``If the professor tells me to make something better, I'll figure out myself what is `better' and how to make it better. But with GPT, I have to say exactly what is better.''}
We believe that this issue could be mitigated by using features as boundary objects and reverting the initiative in feature selection. Instead of a human selecting the feature configuration, the agent could use the features to make confirmations with the human. For example, to generate an ``academic level'' summary, the agent could confirm with the human what ``academic level'' means by asking for confirmations on the complexity level, formality level, and so on. 
Incorporating features as boundary objects presents a promising direction for supporting better human-agent interaction.

\vspace*{-0.1cm}
\paragraph{\textit{\textbf{\footnotesize Issue of Trust}}}
As a professional journalist, C1 had serious trust issues with applying intelligent agents in journalism work: \textit{``Because I know about GPT hallucinations, and if I am saying false things in this material that I produce, that's really bad for me.''}
For C1, the level of trust needed for the agent seems unreachable: \textit{`` (To trust it) I have to check every single thing that it says, and it defeats the purpose of having a tool that increases my efficiency.''}
In the feature-oriented evaluation workflow, we do not consider the evaluation of the trustworthiness of the prompts as it can not be easily measured through any feature that we know of. Supporting the evaluation of trustworthiness remains challenging yet critical for many professional tasks.

\vspace*{-0.1cm}    
% \vspace*{-0.15cm}
\section{Limitations and Future Work}
\sam{
Limitations of the system are centered around features. 
First, the feature metrics have their limitations. For example, the Faithfulness score computation is confined to the token level, which could hinder its accuracy when the entities have different meanings across sentences. The capability of \system \ to capture user intent accurately is thus limited by the feature metrics. We provide a more thorough discussion of the limitations of feature metrics in the supplemental material.
\marginnote{$\triangle$\_8\_1}
}

Developing or selecting suitable features that match the task at hand remains a challenge in the feature-oriented workflow, especially for non-technical prompt designers.
\sam{This includes the challenge to decompose a complicated task into subtasks that are easier to evaluate, which 
remains hard even for technical designers with relevant training~\cite{wu2023crowdsourcingllm}.}
One possible solution is to maintain a repository of features and provide recommendation support for each task, following the approach for facilitating the selection of layouts for large graphs~\cite{kwon2018graphlayout}.
The practitioner review revealed that the evaluation criteria of most prompting tasks could be decomposed into a combination of features, making it possible to cover endlessly long-tail tasks with a finite amount of features. This observation aligns with the findings of Kim et al.~\cite{kim2023evallm}, that task-specific considerations are often secondary to broader criteria.
Technical contributors can conduct meta evaluation on the features to guarantee their efficacy and maintain a repository of the results, thereby benefiting non-technical prompt designers.

Reliance on the dataset hinders generalizability, as one is not always available in every application scenario.
A potential solution is LLM-powered data augmentation~\cite{seo2024retrieval}. Given a small dataset, LLMs have the capability to generate high-quality data points by retrieving and extrapolating from large datasets with sufficient contextual information. The discussion with C3 revealed that data augmentation is a promising direction to deal with the absence of a dataset.

Finally, certain social factors might not be applicable to the feature-oriented workflow, such as trustworthiness, privacy, ethics, and social biases, which have a significant impact on the adoption of LLM-based intelligent agents deeper into our everyday lives. As mentioned in the practitioner review, the level of trust needed for an intelligent agent to conduct certain professional tasks might take a long time to be gained, defeating the purpose of adopting them for efficiency in the first place. This is the same for other social factors, in that intelligent agents can not be recognized without a transition process that lasts for a significant period. We believe evaluating these social factors is critical before deploying prompts for high-stake tasks, but it is out of the scope of our work and we call for future work to continue on this issue.

% \vspace*{-0.15cm}
\vspace*{-0.1cm}    
\section{Conclusion}
Evaluating human-level summarizations of LLMs requires capturing nuanced quality differences, surpassing the capabilities of quality metrics.
Via \system \ , we have shown that feature metrics have the potential to provide actionable insights in the summarization prompt evaluation setting. 
\marginnote{$\triangle$\_9\_1}
While still limited, the practitioner review reveals positive directions to extend the feature-oriented workflow for broader prompting tasks. 
We call for more research to follow this direction to advance prompt evaluation via a human-in-the-loop approach.

%% if specified like this the section will be ommitted in review mode
% \acknowledgments{%
% 	The authors wish to thank the constructive feedback from our anonymous reviewers.
% }
\acknowledgments{%
	This research is supported in part by the UC Climate Action Initiative.
}

\bibliographystyle{abbrv-doi-hyperref}
%\bibliographystyle{abbrv-doi-hyperref-narrow}
%\bibliographystyle{abbrv-doi}
%\bibliographystyle{abbrv-doi-narrow}

% \clearpage

\bibliography{template}

\begin{thebibliography}{10}

\bibitem{openaiprompt}
Prompt engineering from openai.
\newblock \url{https://platform.openai.com/docs/guides/prompt-engineering}.
\newblock Accessed: 2024-02-12.

\bibitem{promptengineeringuide}
Prompt engineering guide.
\newblock \url{https://www.promptingguide.ai/}.
\newblock Accessed: 2024-02-12.

\bibitem{mljar}
Mljar.
\newblock \url{https://mljar.com/}, 2018.
\newblock Accessed: 2024-06-12.

\bibitem{sigopt}
Sigopt.
\newblock \url{https://sigopt.com/}, 2018.
\newblock Accessed: 2024-06-12.

\bibitem{ankerst1999optics}
M.~Ankerst, M.~M. Breunig, H.-P. Kriegel, and J.~Sander.
\newblock Optics: ordering points to identify the clustering structure.
\newblock In {\em SIGMOD '99}, p. 49–60. ACM, New York, 1999. \href{https://doi.org/10.1145/304182.304187}
{doi: {{%
10\hspace{.1pt}\discretionary{.}{%
}{.}\hspace{.4pt}1145\discretionary{/}{%
}{/}304182\hspace{.1pt}\discretionary{.}{%
}{.}\hspace{.4pt}304187}}}


\bibitem{arawjo2023chainforge}
I.~Arawjo, C.~Swoopes, P.~Vaithilingam, M.~Wattenberg, and E.~L. Glassman.
\newblock Chainforge: A visual toolkit for prompt engineering and llm hypothesis testing.
\newblock In {\em CHI '24}. ACM, New York, 2024. \href{https://doi.org/10.1145/3613904.3642016}
{doi: {{%
10\hspace{.1pt}\discretionary{.}{%
}{.}\hspace{.4pt}1145\discretionary{/}{%
}{/}3613904\hspace{.1pt}\discretionary{.}{%
}{.}\hspace{.4pt}3642016}}}


\bibitem{bhandari-etal-2020-evaluating}
M.~Bhandari, P.~N. Gour, A.~Ashfaq, P.~Liu, and G.~Neubig.
\newblock Re-evaluating evaluation in text summarization.
\newblock In B.~Webber, T.~Cohn, Y.~He, and Y.~Liu, eds., {\em EMNLP '20}, pp. 9347--9359. ACL, Online, 2020. \href{https://doi.org/10.18653/v1/2020.emnlp-main.751}
{doi: {{%
10\hspace{.1pt}\discretionary{.}{%
}{.}\hspace{.4pt}18653\discretionary{/}{%
}{/}v1\discretionary{/}{%
}{/}2020\hspace{.1pt}\discretionary{.}{%
}{.}\hspace{.4pt}emnlp\discretionary{%
}{-}{-}main\hspace{.1pt}\discretionary{.}{%
}{.}\hspace{.4pt}751}}}


\bibitem{brade2023promptify}
S.~Brade, B.~Wang, M.~Sousa, S.~Oore, and T.~Grossman.
\newblock Promptify: Text-to-image generation through interactive prompt exploration with large language models.
\newblock In {\em UIST '23}. ACM, New York, 2023. \href{https://doi.org/10.1145/3586183.3606725}
{doi: {{%
10\hspace{.1pt}\discretionary{.}{%
}{.}\hspace{.4pt}1145\discretionary{/}{%
}{/}3586183\hspace{.1pt}\discretionary{.}{%
}{.}\hspace{.4pt}3606725}}}


\bibitem{brown2020language}
T.~Brown, B.~Mann, N.~Ryder, M.~Subbiah, J.~D. Kaplan, P.~Dhariwal, A.~Neelakantan, P.~Shyam, G.~Sastry, A.~Askell, S.~Agarwal, A.~Herbert-Voss, G.~Krueger, T.~Henighan, R.~Child, A.~Ramesh, D.~Ziegler, J.~Wu, C.~Winter, C.~Hesse, M.~Chen, E.~Sigler, M.~Litwin, S.~Gray, B.~Chess, J.~Clark, C.~Berner, S.~McCandlish, A.~Radford, I.~Sutskever, and D.~Amodei.
\newblock Language models are few-shot learners.
\newblock In H.~Larochelle, M.~Ranzato, R.~Hadsell, M.~Balcan, and H.~Lin, eds., {\em Advances in Neural Information Processing Systems}, vol.~33, pp. 1877--1901. Curran Associates, Inc., 2020.

\bibitem{deutsch2022re}
D.~Deutsch, R.~Dror, and D.~Roth.
\newblock Re-examining system-level correlations of automatic summarization evaluation metrics.
\newblock In M.~Carpuat, M.-C. de~Marneffe, and I.~V. Meza~Ruiz, eds., {\em Proceedings of the 2022 Conference of the North American Chapter of the Association for Computational Linguistics: Human Language Technologies}, pp. 6038--6052. ACL, Seattle, 2022. \href{https://doi.org/10.18653/v1/2022.naacl-main.442}
{doi: {{%
10\hspace{.1pt}\discretionary{.}{%
}{.}\hspace{.4pt}18653\discretionary{/}{%
}{/}v1\discretionary{/}{%
}{/}2022\hspace{.1pt}\discretionary{.}{%
}{.}\hspace{.4pt}naacl\discretionary{%
}{-}{-}main\hspace{.1pt}\discretionary{.}{%
}{.}\hspace{.4pt}442}}}


\bibitem{cox2000mds}
J.~{Douglas Carroll} and P.~Arabie.
\newblock Chapter 3 - multidimensional scaling.
\newblock In M.~H. Birnbaum, ed., {\em Measurement, Judgment and Decision Making}, Handbook of Perception and Cognition (Second Edition), pp. 179--250. Academic Press, San Diego, 1998. \href{https://doi.org/10.1016/B978-012099975-0.50005-1}
{doi: {{%
10\hspace{.1pt}\discretionary{.}{%
}{.}\hspace{.4pt}1016\discretionary{/}{%
}{/}B978\discretionary{%
}{-}{-}012099975\discretionary{%
}{-}{-}0\hspace{.1pt}\discretionary{.}{%
}{.}\hspace{.4pt}50005\discretionary{%
}{-}{-}1}}}


\bibitem{durmus-etal-2020-feqa}
E.~Durmus, H.~He, and M.~Diab.
\newblock {FEQA}: A question answering evaluation framework for faithfulness assessment in abstractive summarization.
\newblock In D.~Jurafsky, J.~Chai, N.~Schluter, and J.~Tetreault, eds., {\em Proceedings of the 58th Annual Meeting of the Association for Computational Linguistics}, pp. 5055--5070. ACL, Online, 2020. \href{https://doi.org/10.18653/v1/2020.acl-main.454}
{doi: {{%
10\hspace{.1pt}\discretionary{.}{%
}{.}\hspace{.4pt}18653\discretionary{/}{%
}{/}v1\discretionary{/}{%
}{/}2020\hspace{.1pt}\discretionary{.}{%
}{.}\hspace{.4pt}acl\discretionary{%
}{-}{-}main\hspace{.1pt}\discretionary{.}{%
}{.}\hspace{.4pt}454}}}


\bibitem{durmus-etal-2022-spurious}
E.~Durmus, F.~Ladhak, and T.~Hashimoto.
\newblock Spurious correlations in reference-free evaluation of text generation.
\newblock In S.~Muresan, P.~Nakov, and A.~Villavicencio, eds., {\em Proceedings of the 60th Annual Meeting of the Association for Computational Linguistics (Volume 1: Long Papers)}, pp. 1443--1454. ACL, Dublin, 2022. \href{https://doi.org/10.18653/v1/2022.acl-long.102}
{doi: {{%
10\hspace{.1pt}\discretionary{.}{%
}{.}\hspace{.4pt}18653\discretionary{/}{%
}{/}v1\discretionary{/}{%
}{/}2022\hspace{.1pt}\discretionary{.}{%
}{.}\hspace{.4pt}acl\discretionary{%
}{-}{-}long\hspace{.1pt}\discretionary{.}{%
}{.}\hspace{.4pt}102}}}


\bibitem{fabbri2022qafacteval}
A.~Fabbri, C.-S. Wu, W.~Liu, and C.~Xiong.
\newblock {QAF}act{E}val: Improved {QA}-based factual consistency evaluation for summarization.
\newblock In M.~Carpuat, M.-C. de~Marneffe, and I.~V. Meza~Ruiz, eds., {\em Proceedings of the 2022 Conference of the North American Chapter of the Association for Computational Linguistics: Human Language Technologies}, pp. 2587--2601. ACL, Seattle, 2022. \href{https://doi.org/10.18653/v1/2022.naacl-main.187}
{doi: {{%
10\hspace{.1pt}\discretionary{.}{%
}{.}\hspace{.4pt}18653\discretionary{/}{%
}{/}v1\discretionary{/}{%
}{/}2022\hspace{.1pt}\discretionary{.}{%
}{.}\hspace{.4pt}naacl\discretionary{%
}{-}{-}main\hspace{.1pt}\discretionary{.}{%
}{.}\hspace{.4pt}187}}}


\bibitem{falke-etal-2019-ranking}
T.~Falke, L.~F.~R. Ribeiro, P.~A. Utama, I.~Dagan, and I.~Gurevych.
\newblock Ranking generated summaries by correctness: An interesting but challenging application for natural language inference.
\newblock In A.~Korhonen, D.~Traum, and L.~M{\`a}rquez, eds., {\em Proceedings of the 57th Annual Meeting of the Association for Computational Linguistics}, pp. 2214--2220. ACL, Florence, 2019. \href{https://doi.org/10.18653/v1/P19-1213}
{doi: {{%
10\hspace{.1pt}\discretionary{.}{%
}{.}\hspace{.4pt}18653\discretionary{/}{%
}{/}v1\discretionary{/}{%
}{/}P19\discretionary{%
}{-}{-}1213}}}


\bibitem{golovin2017google}
D.~Golovin, B.~Solnik, S.~Moitra, G.~Kochanski, J.~Karro, and D.~Sculley.
\newblock Google vizier: A service for black-box optimization.
\newblock In {\em KDD '17}, p. 1487–1495. ACM, New York, 2017. \href{https://doi.org/10.1145/3097983.3098043}
{doi: {{%
10\hspace{.1pt}\discretionary{.}{%
}{.}\hspace{.4pt}1145\discretionary{/}{%
}{/}3097983\hspace{.1pt}\discretionary{.}{%
}{.}\hspace{.4pt}3098043}}}


\bibitem{groves2018treat}
I.~Groves, Y.~Tian, and I.~Douratsos.
\newblock Treat the system like a human student: Automatic naturalness evaluation of generated text without reference texts.
\newblock In E.~Krahmer, A.~Gatt, and M.~Goudbeek, eds., {\em Proceedings of the 11th International Conference on Natural Language Generation}, pp. 109--118. ACL, Tilburg University, 2018. \href{https://doi.org/10.18653/v1/W18-6512}
{doi: {{%
10\hspace{.1pt}\discretionary{.}{%
}{.}\hspace{.4pt}18653\discretionary{/}{%
}{/}v1\discretionary{/}{%
}{/}W18\discretionary{%
}{-}{-}6512}}}


\bibitem{heylighen1999formality}
F.~Heylighen and J.-M. Dewaele.
\newblock Formality of language: definition, measurement and behavioral determinants.
\newblock {\em Interner Bericht, Center “Leo Apostel”, Vrije Universiteit Br{\"u}ssel}, 4(1), 1999.

\bibitem{hohman2018visual}
F.~Hohman, M.~Kahng, R.~Pienta, and D.~H. Chau.
\newblock Visual analytics in deep learning: An interrogative survey for the next frontiers.
\newblock {\em IEEE TVCG}, 25(8):2674--2693, 2019. \href{https://doi.org/10.1109/TVCG.2018.2843369}
{doi: {{%
10\hspace{.1pt}\discretionary{.}{%
}{.}\hspace{.4pt}1109\discretionary{/}{%
}{/}TVCG\hspace{.1pt}\discretionary{.}{%
}{.}\hspace{.4pt}2018\hspace{.1pt}\discretionary{.}{%
}{.}\hspace{.4pt}2843369}}}


\bibitem{honnibal2017spacy}
M.~Honnibal and I.~Montani.
\newblock spacy 2: Natural language understanding with bloom embeddings, convolutional neural networks and incremental parsing.
\newblock {\em To appear}, 7(1):411--420, 2017.

\bibitem{huang2024graphimind}
Q.~Huang, M.~Lu, J.~Lanir, D.~Lischinski, D.~Cohen-Or, and H.~Huang.
\newblock Graphimind: Llm-centric interface for information graphics design, 2024. \href{https://doi.org/10.48550/arXiv.2401.13245}
{doi: {{%
10\hspace{.1pt}\discretionary{.}{%
}{.}\hspace{.4pt}48550\discretionary{/}{%
}{/}arXiv\hspace{.1pt}\discretionary{.}{%
}{.}\hspace{.4pt}2401\hspace{.1pt}\discretionary{.}{%
}{.}\hspace{.4pt}13245}}}


\bibitem{hutto2014vader}
C.~Hutto and E.~Gilbert.
\newblock Vader: A parsimonious rule-based model for sentiment analysis of social media text.
\newblock In {\em Proceedings of the international AAAI conference on web and social media}, vol.~8, pp. 216--225, 2014. \href{https://doi.org/10.1609/icwsm.v8i1.14550}
{doi: {{%
10\hspace{.1pt}\discretionary{.}{%
}{.}\hspace{.4pt}1609\discretionary{/}{%
}{/}icwsm\hspace{.1pt}\discretionary{.}{%
}{.}\hspace{.4pt}v8i1\hspace{.1pt}\discretionary{.}{%
}{.}\hspace{.4pt}14550}}}


\bibitem{jiang2022promptmaker}
E.~Jiang, K.~Olson, E.~Toh, A.~Molina, A.~Donsbach, M.~Terry, and C.~J. Cai.
\newblock Promptmaker: Prompt-based prototyping with large language models.
\newblock In {\em CHI EA '22}, CHI EA '22. ACM, New York, 2022. \href{https://doi.org/10.1145/3491101.3503564}
{doi: {{%
10\hspace{.1pt}\discretionary{.}{%
}{.}\hspace{.4pt}1145\discretionary{/}{%
}{/}3491101\hspace{.1pt}\discretionary{.}{%
}{.}\hspace{.4pt}3503564}}}


\bibitem{kim2023cells}
T.~S. Kim, Y.~Lee, M.~Chang, and J.~Kim.
\newblock Cells, generators, and lenses: Design framework for object-oriented interaction with large language models.
\newblock In {\em UIST '23}. ACM, New York, 2023. \href{https://doi.org/10.1145/3586183.3606833}
{doi: {{%
10\hspace{.1pt}\discretionary{.}{%
}{.}\hspace{.4pt}1145\discretionary{/}{%
}{/}3586183\hspace{.1pt}\discretionary{.}{%
}{.}\hspace{.4pt}3606833}}}


\bibitem{kim2023evallm}
T.~S. Kim, Y.~Lee, J.~Shin, Y.-H. Kim, and J.~Kim.
\newblock Evallm: Interactive evaluation of large language model prompts on user-defined criteria.
\newblock In {\em CHI '24}. ACM, New York, 2024. \href{https://doi.org/10.1145/3613904.3642216}
{doi: {{%
10\hspace{.1pt}\discretionary{.}{%
}{.}\hspace{.4pt}1145\discretionary{/}{%
}{/}3613904\hspace{.1pt}\discretionary{.}{%
}{.}\hspace{.4pt}3642216}}}


\bibitem{kincaid1975derivation}
J.~P. Kincaid, R.~P. Fishburne~Jr, R.~L. Rogers, and B.~S. Chissom.
\newblock Derivation of new readability formulas (automated readability index, fog count and flesch reading ease formula) for navy enlisted personnel.
\newblock 1975. \href{https://doi.org/10.21236/ada006655}
{doi: {{%
10\hspace{.1pt}\discretionary{.}{%
}{.}\hspace{.4pt}21236\discretionary{/}{%
}{/}ada006655}}}


\bibitem{kwon2018graphlayout}
O.-H. Kwon, T.~Crnovrsanin, and K.-L. Ma.
\newblock What would a graph look like in this layout? a machine learning approach to large graph visualization.
\newblock {\em IEEE TVCG}, 24(1):478--488, 2018. \href{https://doi.org/10.1109/TVCG.2017.2743858}
{doi: {{%
10\hspace{.1pt}\discretionary{.}{%
}{.}\hspace{.4pt}1109\discretionary{/}{%
}{/}TVCG\hspace{.1pt}\discretionary{.}{%
}{.}\hspace{.4pt}2017\hspace{.1pt}\discretionary{.}{%
}{.}\hspace{.4pt}2743858}}}


\bibitem{laban2022summac}
P.~Laban, T.~Schnabel, P.~N. Bennett, and M.~A. Hearst.
\newblock {S}umma{C}: Re-visiting {NLI}-based models for inconsistency detection in summarization.
\newblock {\em Transactions of the Association for Computational Linguistics}, 10:163--177, 2022. \href{https://doi.org/10.1162/tacl_a_00453}
{doi: {{%
10\hspace{.1pt}\discretionary{.}{%
}{.}\hspace{.4pt}1162\discretionary{/}{%
}{/}tacl\_a\_00453}}}


\bibitem{li2023evaluateChatgpt}
B.~Li, G.~Fang, Y.~Yang, Q.~Wang, W.~Ye, W.~Zhao, and S.~Zhang.
\newblock Evaluating chatgpt's information extraction capabilities: An assessment of performance, explainability, calibration, and faithfulness, 2023. \href{https://doi.org/10.48550/arXiv.2304.11633}
{doi: {{%
10\hspace{.1pt}\discretionary{.}{%
}{.}\hspace{.4pt}48550\discretionary{/}{%
}{/}arXiv\hspace{.1pt}\discretionary{.}{%
}{.}\hspace{.4pt}2304\hspace{.1pt}\discretionary{.}{%
}{.}\hspace{.4pt}11633}}}


\bibitem{lin2004rouge}
C.-Y. Lin.
\newblock {ROUGE}: A package for automatic evaluation of summaries.
\newblock In {\em Text Summarization Branches Out}, pp. 74--81. ACL, Barcelona, 2004.

\bibitem{liu2024cliqueparcel}
J.~Liu, T.~Yang, and J.~Neville.
\newblock Cliqueparcel: An approach for batching llm prompts that jointly optimizes efficiency and faithfulness, 2024. \href{https://doi.org/10.48550/arXiv.2402.14833}
{doi: {{%
10\hspace{.1pt}\discretionary{.}{%
}{.}\hspace{.4pt}48550\discretionary{/}{%
}{/}arXiv\hspace{.1pt}\discretionary{.}{%
}{.}\hspace{.4pt}2402\hspace{.1pt}\discretionary{.}{%
}{.}\hspace{.4pt}14833}}}


\bibitem{liu2018analyzetraining}
M.~Liu, J.~Shi, K.~Cao, J.~Zhu, and S.~Liu.
\newblock Analyzing the training processes of deep generative models.
\newblock {\em IEEE TVCG}, 24(1):77--87, 2018. \href{https://doi.org/10.1109/TVCG.2017.2744938}
{doi: {{%
10\hspace{.1pt}\discretionary{.}{%
}{.}\hspace{.4pt}1109\discretionary{/}{%
}{/}TVCG\hspace{.1pt}\discretionary{.}{%
}{.}\hspace{.4pt}2017\hspace{.1pt}\discretionary{.}{%
}{.}\hspace{.4pt}2744938}}}


\bibitem{lloyd1982kmeans}
S.~Lloyd.
\newblock Least squares quantization in pcm.
\newblock {\em IEEE TIT}, 28(2):129--137, 1982. \href{https://doi.org/10.1109/TIT.1982.1056489}
{doi: {{%
10\hspace{.1pt}\discretionary{.}{%
}{.}\hspace{.4pt}1109\discretionary{/}{%
}{/}TIT\hspace{.1pt}\discretionary{.}{%
}{.}\hspace{.4pt}1982\hspace{.1pt}\discretionary{.}{%
}{.}\hspace{.4pt}1056489}}}


\bibitem{lu-etal-2022-fantastically}
Y.~Lu, M.~Bartolo, A.~Moore, S.~Riedel, and P.~Stenetorp.
\newblock Fantastically ordered prompts and where to find them: Overcoming few-shot prompt order sensitivity.
\newblock In S.~Muresan, P.~Nakov, and A.~Villavicencio, eds., {\em Proceedings of the 60th Annual Meeting of the Association for Computational Linguistics (Volume 1: Long Papers)}, pp. 8086--8098. Association for Computational Linguistics, Dublin, Ireland, May 2022. \href{https://doi.org/10.18653/v1/2022.acl-long.556}
{doi: {{%
10\hspace{.1pt}\discretionary{.}{%
}{.}\hspace{.4pt}18653\discretionary{/}{%
}{/}v1\discretionary{/}{%
}{/}2022\hspace{.1pt}\discretionary{.}{%
}{.}\hspace{.4pt}acl\discretionary{%
}{-}{-}long\hspace{.1pt}\discretionary{.}{%
}{.}\hspace{.4pt}556}}}


\bibitem{macneil2023promptmiddleware}
S.~MacNeil, A.~Tran, J.~Kim, Z.~Huang, S.~Bernstein, and D.~Mogil.
\newblock Prompt middleware: Mapping prompts for large language models to ui affordances, 2023. \href{https://doi.org/10.48550/arXiv.2307.01142}
{doi: {{%
10\hspace{.1pt}\discretionary{.}{%
}{.}\hspace{.4pt}48550\discretionary{/}{%
}{/}arXiv\hspace{.1pt}\discretionary{.}{%
}{.}\hspace{.4pt}2307\hspace{.1pt}\discretionary{.}{%
}{.}\hspace{.4pt}01142}}}


\bibitem{maynez-etal-2020-faithfulness}
J.~Maynez, S.~Narayan, B.~Bohnet, and R.~McDonald.
\newblock On faithfulness and factuality in abstractive summarization.
\newblock In D.~Jurafsky, J.~Chai, N.~Schluter, and J.~Tetreault, eds., {\em Proceedings of the 58th Annual Meeting of the Association for Computational Linguistics}, pp. 1906--1919. ACL, Online, 2020. \href{https://doi.org/10.18653/v1/2020.acl-main.173}
{doi: {{%
10\hspace{.1pt}\discretionary{.}{%
}{.}\hspace{.4pt}18653\discretionary{/}{%
}{/}v1\discretionary{/}{%
}{/}2020\hspace{.1pt}\discretionary{.}{%
}{.}\hspace{.4pt}acl\discretionary{%
}{-}{-}main\hspace{.1pt}\discretionary{.}{%
}{.}\hspace{.4pt}173}}}


\bibitem{mccarthy2010mtld}
P.~M. McCarthy and S.~Jarvis.
\newblock Mtld, vocd-d, and hd-d: A validation study of sophisticated approaches to lexical diversity assessment.
\newblock {\em Behavior research methods}, 42(2):381--392, 2010. \href{https://doi.org/10.3758/BRM.42.2.381}
{doi: {{%
10\hspace{.1pt}\discretionary{.}{%
}{.}\hspace{.4pt}3758\discretionary{/}{%
}{/}BRM\hspace{.1pt}\discretionary{.}{%
}{.}\hspace{.4pt}42\hspace{.1pt}\discretionary{.}{%
}{.}\hspace{.4pt}2\hspace{.1pt}\discretionary{.}{%
}{.}\hspace{.4pt}381}}}


\bibitem{mcinnes2020umap}
L.~McInnes, J.~Healy, and J.~Melville.
\newblock Umap: Uniform manifold approximation and projection for dimension reduction, 2020. \href{https://doi.org/10.48550/arXiv.1802.03426}
{doi: {{%
10\hspace{.1pt}\discretionary{.}{%
}{.}\hspace{.4pt}48550\discretionary{/}{%
}{/}arXiv\hspace{.1pt}\discretionary{.}{%
}{.}\hspace{.4pt}1802\hspace{.1pt}\discretionary{.}{%
}{.}\hspace{.4pt}03426}}}


\bibitem{mishra2023promptaid}
A.~Mishra, U.~Soni, A.~Arunkumar, J.~Huang, B.~C. Kwon, and C.~Bryan.
\newblock Promptaid: Prompt exploration, perturbation, testing and iteration using visual analytics for large language models, 2023. \href{https://doi.org/10.48550/arXiv.2304.01964}
{doi: {{%
10\hspace{.1pt}\discretionary{.}{%
}{.}\hspace{.4pt}48550\discretionary{/}{%
}{/}arXiv\hspace{.1pt}\discretionary{.}{%
}{.}\hspace{.4pt}2304\hspace{.1pt}\discretionary{.}{%
}{.}\hspace{.4pt}01964}}}


\bibitem{munoz2023contrasting}
A.~Muñoz-Ortiz, C.~Gómez-Rodríguez, and D.~Vilares.
\newblock Contrasting linguistic patterns in human and llm-generated text, 2023. \href{https://doi.org/10.48550/arXiv.2308.09067}
{doi: {{%
10\hspace{.1pt}\discretionary{.}{%
}{.}\hspace{.4pt}48550\discretionary{/}{%
}{/}arXiv\hspace{.1pt}\discretionary{.}{%
}{.}\hspace{.4pt}2308\hspace{.1pt}\discretionary{.}{%
}{.}\hspace{.4pt}09067}}}


\bibitem{novikova-etal-2017-need}
J.~Novikova, O.~Du{\v{s}}ek, A.~Cercas~Curry, and V.~Rieser.
\newblock Why we need new evaluation metrics for {NLG}.
\newblock In M.~Palmer, R.~Hwa, and S.~Riedel, eds., {\em EMNLP '27}, pp. 2241--2252. ACL, Copenhagen, 2017. \href{https://doi.org/10.18653/v1/D17-1238}
{doi: {{%
10\hspace{.1pt}\discretionary{.}{%
}{.}\hspace{.4pt}18653\discretionary{/}{%
}{/}v1\discretionary{/}{%
}{/}D17\discretionary{%
}{-}{-}1238}}}


\bibitem{ouyang2023llm}
S.~Ouyang, J.~M. Zhang, M.~Harman, and M.~Wang.
\newblock Llm is like a box of chocolates: the non-determinism of chatgpt in code generation.
\newblock {\em arXiv preprint arXiv:2308.02828}, 2023.

\bibitem{palubinskas2017image}
G.~Palubinskas.
\newblock Image similarity/distance measures: what is really behind mse and ssim?
\newblock {\em International Journal of Image and Data Fusion}, 8(1):32--53, 2017. \href{https://doi.org/10.1080/19479832.2016.1273259}
{doi: {{%
10\hspace{.1pt}\discretionary{.}{%
}{.}\hspace{.4pt}1080\discretionary{/}{%
}{/}19479832\hspace{.1pt}\discretionary{.}{%
}{.}\hspace{.4pt}2016\hspace{.1pt}\discretionary{.}{%
}{.}\hspace{.4pt}1273259}}}


\bibitem{papineni2002bleu}
K.~Papineni, S.~Roukos, T.~Ward, and W.-J. Zhu.
\newblock {B}leu: a method for automatic evaluation of machine translation.
\newblock In P.~Isabelle, E.~Charniak, and D.~Lin, eds., {\em Proceedings of the 40th Annual Meeting of the Association for Computational Linguistics}, pp. 311--318. ACL, Philadelphia, 2002. \href{https://doi.org/10.3115/1073083.1073135}
{doi: {{%
10\hspace{.1pt}\discretionary{.}{%
}{.}\hspace{.4pt}3115\discretionary{/}{%
}{/}1073083\hspace{.1pt}\discretionary{.}{%
}{.}\hspace{.4pt}1073135}}}


\bibitem{petridis2023anglekindling}
S.~Petridis, N.~Diakopoulos, K.~Crowston, M.~Hansen, K.~Henderson, S.~Jastrzebski, J.~V. Nickerson, and L.~B. Chilton.
\newblock Anglekindling: Supporting journalistic angle ideation with large language models.
\newblock In {\em CHI '23}. ACM, New York, 2023. \href{https://doi.org/10.1145/3544548.3580907}
{doi: {{%
10\hspace{.1pt}\discretionary{.}{%
}{.}\hspace{.4pt}1145\discretionary{/}{%
}{/}3544548\hspace{.1pt}\discretionary{.}{%
}{.}\hspace{.4pt}3580907}}}


\bibitem{pezzotti2018deepeyes}
N.~Pezzotti, T.~Höllt, J.~Van~Gemert, B.~P. Lelieveldt, E.~Eisemann, and A.~Vilanova.
\newblock Deepeyes: Progressive visual analytics for designing deep neural networks.
\newblock {\em IEEE TVCG}, 24(1):98--108, 2018. \href{https://doi.org/10.1109/TVCG.2017.2744358}
{doi: {{%
10\hspace{.1pt}\discretionary{.}{%
}{.}\hspace{.4pt}1109\discretionary{/}{%
}{/}TVCG\hspace{.1pt}\discretionary{.}{%
}{.}\hspace{.4pt}2017\hspace{.1pt}\discretionary{.}{%
}{.}\hspace{.4pt}2744358}}}


\bibitem{pu-demberg-2023-chatgpt}
D.~Pu and V.~Demberg.
\newblock {C}hat{GPT} vs human-authored text: Insights into controllable text summarization and sentence style transfer.
\newblock In V.~Padmakumar, G.~Vallejo, and Y.~Fu, eds., {\em Proceedings of the 61st Annual Meeting of the Association for Computational Linguistics (Volume 4: Student Research Workshop)}, pp. 1--18. ACL, Toronto, 2023. \href{https://doi.org/10.18653/v1/2023.acl-srw.1}
{doi: {{%
10\hspace{.1pt}\discretionary{.}{%
}{.}\hspace{.4pt}18653\discretionary{/}{%
}{/}v1\discretionary{/}{%
}{/}2023\hspace{.1pt}\discretionary{.}{%
}{.}\hspace{.4pt}acl\discretionary{%
}{-}{-}srw\hspace{.1pt}\discretionary{.}{%
}{.}\hspace{.4pt}1}}}


\bibitem{scheepens2016animatedflow}
R.~Scheepens, C.~Hurter, H.~Van De~Wetering, and J.~J. Van~Wijk.
\newblock Visualization, selection, and analysis of traffic flows.
\newblock {\em IEEE TVCG}, 22(1):379--388, 2016. \href{https://doi.org/10.1109/TVCG.2015.2467112}
{doi: {{%
10\hspace{.1pt}\discretionary{.}{%
}{.}\hspace{.4pt}1109\discretionary{/}{%
}{/}TVCG\hspace{.1pt}\discretionary{.}{%
}{.}\hspace{.4pt}2015\hspace{.1pt}\discretionary{.}{%
}{.}\hspace{.4pt}2467112}}}


\bibitem{scholkopf1997kernel}
B.~Sch{\"o}lkopf, A.~Smola, and K.-R. M{\"u}ller.
\newblock Kernel principal component analysis.
\newblock In W.~Gerstner, A.~Germond, M.~Hasler, and J.-D. Nicoud, eds., {\em Artificial Neural Networks --- ICANN'97}, pp. 583--588. Springer Berlin Heidelberg, Berlin, 1997. \href{https://doi.org/10.1007/BFb0020217}
{doi: {{%
10\hspace{.1pt}\discretionary{.}{%
}{.}\hspace{.4pt}1007\discretionary{/}{%
}{/}BFb0020217}}}


\bibitem{scialom-etal-2021-questeval}
T.~Scialom, P.-A. Dray, S.~Lamprier, B.~Piwowarski, J.~Staiano, A.~Wang, and P.~Gallinari.
\newblock {Q}uest{E}val: Summarization asks for fact-based evaluation.
\newblock In M.-F. Moens, X.~Huang, L.~Specia, and S.~W.-t. Yih, eds., {\em EMNLP '21}, pp. 6594--6604. ACL, Online and Punta Cana, 2021. \href{https://doi.org/10.18653/v1/2021.emnlp-main.529}
{doi: {{%
10\hspace{.1pt}\discretionary{.}{%
}{.}\hspace{.4pt}18653\discretionary{/}{%
}{/}v1\discretionary{/}{%
}{/}2021\hspace{.1pt}\discretionary{.}{%
}{.}\hspace{.4pt}emnlp\discretionary{%
}{-}{-}main\hspace{.1pt}\discretionary{.}{%
}{.}\hspace{.4pt}529}}}


\bibitem{seo2024retrieval}
M.~Seo, J.~Baek, J.~Thorne, and S.~J. Hwang.
\newblock Retrieval-augmented data augmentation for low-resource domain tasks, 2024. \href{https://doi.org/10.48550/arXiv.2402.13482}
{doi: {{%
10\hspace{.1pt}\discretionary{.}{%
}{.}\hspace{.4pt}48550\discretionary{/}{%
}{/}arXiv\hspace{.1pt}\discretionary{.}{%
}{.}\hspace{.4pt}2402\hspace{.1pt}\discretionary{.}{%
}{.}\hspace{.4pt}13482}}}


\bibitem{shen2023large}
C.~Shen, L.~Cheng, X.-P. Nguyen, Y.~You, and L.~Bing.
\newblock Large language models are not yet human-level evaluators for abstractive summarization.
\newblock In H.~Bouamor, J.~Pino, and K.~Bali, eds., {\em Findings of the Association for Computational Linguistics: EMNLP 2023}, pp. 4215--4233. ACL, Singapore, 2023. \href{https://doi.org/10.18653/v1/2023.findings-emnlp.278}
{doi: {{%
10\hspace{.1pt}\discretionary{.}{%
}{.}\hspace{.4pt}18653\discretionary{/}{%
}{/}v1\discretionary{/}{%
}{/}2023\hspace{.1pt}\discretionary{.}{%
}{.}\hspace{.4pt}findings\discretionary{%
}{-}{-}emnlp\hspace{.1pt}\discretionary{.}{%
}{.}\hspace{.4pt}278}}}


\bibitem{srinivasan2018barcomparison}
A.~Srinivasan, M.~Brehmer, B.~Lee, and S.~M. Drucker.
\newblock What's the difference? evaluating variations of multi-series bar charts for visual comparison tasks.
\newblock In {\em CHI '18}, p. 1–12. ACM, New York, 2018. \href{https://doi.org/10.1145/3173574.3173878}
{doi: {{%
10\hspace{.1pt}\discretionary{.}{%
}{.}\hspace{.4pt}1145\discretionary{/}{%
}{/}3173574\hspace{.1pt}\discretionary{.}{%
}{.}\hspace{.4pt}3173878}}}


\bibitem{star1989institutional}
S.~L. Star and J.~R. Griesemer.
\newblock Institutional ecology, `translations' and boundary objects: Amateurs and professionals in berkeley's museum of vertebrate zoology, 1907-39.
\newblock {\em Social Studies of Science}, 19(3):387--420, 1989. \href{https://doi.org/10.1177/030631289019003001}
{doi: {{%
10\hspace{.1pt}\discretionary{.}{%
}{.}\hspace{.4pt}1177\discretionary{/}{%
}{/}030631289019003001}}}


\bibitem{strathern1997improving}
M.~Strathern.
\newblock ‘improving ratings’: audit in the british university system.
\newblock {\em European Review}, 5(3):305–321, 1997. \href{https://doi.org/10.1017/s1062798700002660}
{doi: {{%
10\hspace{.1pt}\discretionary{.}{%
}{.}\hspace{.4pt}1017\discretionary{/}{%
}{/}s1062798700002660}}}


\bibitem{strobelt2018lstmvis}
H.~Strobelt, S.~Gehrmann, H.~Pfister, and A.~M. Rush.
\newblock Lstmvis: A tool for visual analysis of hidden state dynamics in recurrent neural networks.
\newblock {\em IEEE TVCG}, 24(1):667--676, 2018. \href{https://doi.org/10.1109/TVCG.2017.2744158}
{doi: {{%
10\hspace{.1pt}\discretionary{.}{%
}{.}\hspace{.4pt}1109\discretionary{/}{%
}{/}TVCG\hspace{.1pt}\discretionary{.}{%
}{.}\hspace{.4pt}2017\hspace{.1pt}\discretionary{.}{%
}{.}\hspace{.4pt}2744158}}}


\bibitem{strobelt2022promptide}
H.~Strobelt, A.~Webson, V.~Sanh, B.~Hoover, J.~Beyer, H.~Pfister, and A.~M. Rush.
\newblock Interactive and visual prompt engineering for ad-hoc task adaptation with large language models.
\newblock {\em IEEE TVCG}, 29(01):1146--1156, 2023. \href{https://doi.org/10.1109/TVCG.2022.3209479}
{doi: {{%
10\hspace{.1pt}\discretionary{.}{%
}{.}\hspace{.4pt}1109\discretionary{/}{%
}{/}TVCG\hspace{.1pt}\discretionary{.}{%
}{.}\hspace{.4pt}2022\hspace{.1pt}\discretionary{.}{%
}{.}\hspace{.4pt}3209479}}}


\bibitem{suvcik2023prompterator}
S.~Su{\v{c}}ik, D.~Skala, A.~{\v{S}}vec, P.~Hra{\v{s}}ka, and M.~{\v{S}}uppa.
\newblock Prompterator: Iterate efficiently towards more effective prompts.
\newblock In Y.~Feng and E.~Lefever, eds., {\em Proceedings of the 2023 Conference on Empirical Methods in Natural Language Processing: System Demonstrations}, pp. 471--478. ACL, Singapore, 2023. \href{https://doi.org/10.18653/v1/2023.emnlp-demo.43}
{doi: {{%
10\hspace{.1pt}\discretionary{.}{%
}{.}\hspace{.4pt}18653\discretionary{/}{%
}{/}v1\discretionary{/}{%
}{/}2023\hspace{.1pt}\discretionary{.}{%
}{.}\hspace{.4pt}emnlp\discretionary{%
}{-}{-}demo\hspace{.1pt}\discretionary{.}{%
}{.}\hspace{.4pt}43}}}


\bibitem{temperley2008dependency}
D.~Temperley.
\newblock Dependency-length minimization in natural and artificial languages∗.
\newblock {\em Journal of Quantitative Linguistics}, 15(3):256--282, 2008. \href{https://doi.org/10.1080/09296170802159512}
{doi: {{%
10\hspace{.1pt}\discretionary{.}{%
}{.}\hspace{.4pt}1080\discretionary{/}{%
}{/}09296170802159512}}}


\bibitem{tjuatja2023llms}
L.~Tjuatja, V.~Chen, S.~T. Wu, A.~Talwalkar, and G.~Neubig.
\newblock Do llms exhibit human-like response biases? a case study in survey design, 2024. \href{https://doi.org/10.48550/arXiv.2311.04076}
{doi: {{%
10\hspace{.1pt}\discretionary{.}{%
}{.}\hspace{.4pt}48550\discretionary{/}{%
}{/}arXiv\hspace{.1pt}\discretionary{.}{%
}{.}\hspace{.4pt}2311\hspace{.1pt}\discretionary{.}{%
}{.}\hspace{.4pt}04076}}}


\bibitem{van2008tsne}
L.~van~der Maaten and G.~Hinton.
\newblock Visualizing data using t-sne.
\newblock {\em Journal of Machine Learning Research}, 9(86):2579--2605, 2008.

\bibitem{wang2023dataformulator}
C.~Wang, J.~Thompson, and B.~Lee.
\newblock Data formulator: Ai-powered concept-driven visualization authoring.
\newblock {\em IEEE TVCG}, 30(1):1128--1138, 2024. \href{https://doi.org/10.1109/TVCG.2023.3326585}
{doi: {{%
10\hspace{.1pt}\discretionary{.}{%
}{.}\hspace{.4pt}1109\discretionary{/}{%
}{/}TVCG\hspace{.1pt}\discretionary{.}{%
}{.}\hspace{.4pt}2023\hspace{.1pt}\discretionary{.}{%
}{.}\hspace{.4pt}3326585}}}


\bibitem{wang2023positionbias}
P.~Wang, L.~Li, L.~Chen, Z.~Cai, D.~Zhu, B.~Lin, Y.~Cao, Q.~Liu, T.~Liu, and Z.~Sui.
\newblock Large language models are not fair evaluators, 2023. \href{https://doi.org/10.48550/arXiv.2305.17926}
{doi: {{%
10\hspace{.1pt}\discretionary{.}{%
}{.}\hspace{.4pt}48550\discretionary{/}{%
}{/}arXiv\hspace{.1pt}\discretionary{.}{%
}{.}\hspace{.4pt}2305\hspace{.1pt}\discretionary{.}{%
}{.}\hspace{.4pt}17926}}}


\bibitem{wang2019atmseer}
Q.~Wang, Y.~Ming, Z.~Jin, Q.~Shen, D.~Liu, M.~J. Smith, K.~Veeramachaneni, and H.~Qu.
\newblock Atmseer: Increasing transparency and controllability in automated machine learning.
\newblock In {\em CHI '19}, p. 1–12. ACM, New York, 2019. \href{https://doi.org/10.1145/3290605.3300911}
{doi: {{%
10\hspace{.1pt}\discretionary{.}{%
}{.}\hspace{.4pt}1145\discretionary{/}{%
}{/}3290605\hspace{.1pt}\discretionary{.}{%
}{.}\hspace{.4pt}3300911}}}


\bibitem{wang2023pandalm}
Y.~Wang, Z.~Yu, Z.~Zeng, L.~Yang, C.~Wang, H.~Chen, C.~Jiang, R.~Xie, J.~Wang, X.~Xie, W.~Ye, S.-B. Zhang, and Y.~Zhang.
\newblock Pandalm: An automatic evaluation benchmark for llm instruction tuning optimization.
\newblock In {\em ICLR 2024}, 2023. \href{https://doi.org/10.48550/arXiv.2306.05087}
{doi: {{%
10\hspace{.1pt}\discretionary{.}{%
}{.}\hspace{.4pt}48550\discretionary{/}{%
}{/}arXiv\hspace{.1pt}\discretionary{.}{%
}{.}\hspace{.4pt}2306\hspace{.1pt}\discretionary{.}{%
}{.}\hspace{.4pt}05087}}}


\bibitem{wei2022chain}
J.~Wei, X.~Wang, D.~Schuurmans, M.~Bosma, B.~Ichter, F.~Xia, E.~Chi, Q.~Le, and D.~Zhou.
\newblock Chain-of-thought prompting elicits reasoning in large language models, 2023. \href{https://doi.org/10.48550/arXiv.2201.11903}
{doi: {{%
10\hspace{.1pt}\discretionary{.}{%
}{.}\hspace{.4pt}48550\discretionary{/}{%
}{/}arXiv\hspace{.1pt}\discretionary{.}{%
}{.}\hspace{.4pt}2201\hspace{.1pt}\discretionary{.}{%
}{.}\hspace{.4pt}11903}}}


\bibitem{wongsuphasawat2017visualizing}
K.~Wongsuphasawat, D.~Smilkov, J.~Wexler, J.~Wilson, D.~Mané, D.~Fritz, D.~Krishnan, F.~B. Viégas, and M.~Wattenberg.
\newblock Visualizing dataflow graphs of deep learning models in tensorflow.
\newblock {\em IEEE TVCG}, 24(1):1--12, 2018. \href{https://doi.org/10.1109/TVCG.2017.2744878}
{doi: {{%
10\hspace{.1pt}\discretionary{.}{%
}{.}\hspace{.4pt}1109\discretionary{/}{%
}{/}TVCG\hspace{.1pt}\discretionary{.}{%
}{.}\hspace{.4pt}2017\hspace{.1pt}\discretionary{.}{%
}{.}\hspace{.4pt}2744878}}}


\bibitem{wu2023crowdsourcingllm}
T.~Wu, H.~Zhu, M.~Albayrak, A.~Axon, A.~Bertsch, W.~Deng, Z.~Ding, B.~Guo, S.~Gururaja, T.-S. Kuo, J.~T. Liang, R.~Liu, I.~Mandal, J.~Milbauer, X.~Ni, N.~Padmanabhan, S.~Ramkumar, A.~Sudjianto, J.~Taylor, Y.-J. Tseng, P.~Vaidos, Z.~Wu, W.~Wu, and C.~Yang.
\newblock Llms as workers in human-computational algorithms? replicating crowdsourcing pipelines with llms, 2023. \href{https://doi.org/10.48550/arXiv.2307.10168}
{doi: {{%
10\hspace{.1pt}\discretionary{.}{%
}{.}\hspace{.4pt}48550\discretionary{/}{%
}{/}arXiv\hspace{.1pt}\discretionary{.}{%
}{.}\hspace{.4pt}2307\hspace{.1pt}\discretionary{.}{%
}{.}\hspace{.4pt}10168}}}


\bibitem{ye2024genai}
Y.~Ye, J.~Hao, Y.~Hou, Z.~Wang, S.~Xiao, Y.~Luo, and W.~Zeng.
\newblock Generative ai for visualization: State of the art and future directions.
\newblock {\em Visual Informatics}, 8(2):43--66, 2024. \href{https://doi.org/10.1016/j.visinf.2024.04.003}
{doi: {{%
10\hspace{.1pt}\discretionary{.}{%
}{.}\hspace{.4pt}1016\discretionary{/}{%
}{/}j\hspace{.1pt}\discretionary{.}{%
}{.}\hspace{.4pt}visinf\hspace{.1pt}\discretionary{.}{%
}{.}\hspace{.4pt}2024\hspace{.1pt}\discretionary{.}{%
}{.}\hspace{.4pt}04\hspace{.1pt}\discretionary{.}{%
}{.}\hspace{.4pt}003}}}


\bibitem{yen2009clustersampling}
S.-J. Yen and Y.-S. Lee.
\newblock Cluster-based under-sampling approaches for imbalanced data distributions.
\newblock {\em Expert Systems with Applications}, 36(3, Part 1):5718--5727, 2009. \href{https://doi.org/10.1016/j.eswa.2008.06.108}
{doi: {{%
10\hspace{.1pt}\discretionary{.}{%
}{.}\hspace{.4pt}1016\discretionary{/}{%
}{/}j\hspace{.1pt}\discretionary{.}{%
}{.}\hspace{.4pt}eswa\hspace{.1pt}\discretionary{.}{%
}{.}\hspace{.4pt}2008\hspace{.1pt}\discretionary{.}{%
}{.}\hspace{.4pt}06\hspace{.1pt}\discretionary{.}{%
}{.}\hspace{.4pt}108}}}


\bibitem{zamfirescu2023johnny}
J.~Zamfirescu-Pereira, R.~Y. Wong, B.~Hartmann, and Q.~Yang.
\newblock Why johnny can’t prompt: How non-ai experts try (and fail) to design llm prompts.
\newblock In {\em CHI '23}. ACM, New York, 2023. \href{https://doi.org/10.1145/3544548.3581388}
{doi: {{%
10\hspace{.1pt}\discretionary{.}{%
}{.}\hspace{.4pt}1145\discretionary{/}{%
}{/}3544548\hspace{.1pt}\discretionary{.}{%
}{.}\hspace{.4pt}3581388}}}


\bibitem{Zhang2020BERTScore}
T.~Zhang, V.~Kishore, F.~Wu, K.~Q. Weinberger, and Y.~Artzi.
\newblock Bertscore: Evaluating text generation with bert, 2020. \href{https://doi.org/10.48550/arXiv.1904.09675}
{doi: {{%
10\hspace{.1pt}\discretionary{.}{%
}{.}\hspace{.4pt}48550\discretionary{/}{%
}{/}arXiv\hspace{.1pt}\discretionary{.}{%
}{.}\hspace{.4pt}1904\hspace{.1pt}\discretionary{.}{%
}{.}\hspace{.4pt}09675}}}


\bibitem{zhang2301benchmarking}
T.~Zhang, F.~Ladhak, E.~Durmus, P.~Liang, K.~McKeown, and T.~B. Hashimoto.
\newblock {Benchmarking Large Language Models for News Summarization}.
\newblock {\em Transactions of the Association for Computational Linguistics}, 12:39--57, 2024. \href{https://doi.org/10.1162/tacl_a_00632}
{doi: {{%
10\hspace{.1pt}\discretionary{.}{%
}{.}\hspace{.4pt}1162\discretionary{/}{%
}{/}tacl\_a\_00632}}}


\bibitem{zhang2024benchmarking}
T.~Zhang, F.~Ladhak, E.~Durmus, P.~Liang, K.~McKeown, and T.~B. Hashimoto.
\newblock {Benchmarking Large Language Models for News Summarization}.
\newblock {\em Transactions of the Association for Computational Linguistics}, 12:39--57, 2024. \href{https://doi.org/10.1162/tacl_a_00632}
{doi: {{%
10\hspace{.1pt}\discretionary{.}{%
}{.}\hspace{.4pt}1162\discretionary{/}{%
}{/}tacl\_a\_00632}}}


\bibitem{zhao-etal-2019-moverscore}
W.~Zhao, M.~Peyrard, F.~Liu, Y.~Gao, C.~M. Meyer, and S.~Eger.
\newblock {M}over{S}core: Text generation evaluating with contextualized embeddings and earth mover distance.
\newblock In K.~Inui, J.~Jiang, V.~Ng, and X.~Wan, eds., {\em EMNLP-IJCNLP}, pp. 563--578. ACL, Hong Kong, 2019. \href{https://doi.org/10.18653/v1/D19-1053}
{doi: {{%
10\hspace{.1pt}\discretionary{.}{%
}{.}\hspace{.4pt}18653\discretionary{/}{%
}{/}v1\discretionary{/}{%
}{/}D19\discretionary{%
}{-}{-}1053}}}


\bibitem{zheng2023llmjudge}
L.~Zheng, W.-L. Chiang, Y.~Sheng, S.~Zhuang, Z.~Wu, Y.~Zhuang, Z.~Lin, Z.~Li, D.~Li, E.~Xing, H.~Zhang, J.~E. Gonzalez, and I.~Stoica.
\newblock Judging llm-as-a-judge with mt-bench and chatbot arena.
\newblock In A.~Oh, T.~Neumann, A.~Globerson, K.~Saenko, M.~Hardt, and S.~Levine, eds., {\em Advances in Neural Information Processing Systems}, vol.~36, pp. 46595--46623. Curran Associates, Inc., 2023. \href{https://doi.org/10.48550/arXiv.2306.05685}
{doi: {{%
10\hspace{.1pt}\discretionary{.}{%
}{.}\hspace{.4pt}48550\discretionary{/}{%
}{/}arXiv\hspace{.1pt}\discretionary{.}{%
}{.}\hspace{.4pt}2306\hspace{.1pt}\discretionary{.}{%
}{.}\hspace{.4pt}05685}}}


\end{thebibliography}

% \section{Troubleshooting}
% \label{appendix:troubleshooting}

% \subsection{ifpdf error}

% If you receive compilation errors along the lines of \texttt{Package ifpdf Error: Name clash, \textbackslash ifpdf is already defined} then please add a new line \verb|\let\ifpdf\relax| right after the \verb|\documentclass[journal]{vgtc}| call.
% Note that your error is due to packages you use that define \verb|\ifpdf| which is obsolete (the result is that \verb|\ifpdf| is defined twice); these packages should be changed to use \verb|ifpdf| package instead.

% \subsection{\texttt{pdfendlink} error}

% Occasionally (for some \LaTeX\ distributions) this hyper-linked bib\TeX\ style may lead to \textbf{compilation errors} (\texttt{pdfendlink ended up in different nesting level ...}) if a reference entry is broken across two pages (due to a bug in \verb|hyperref|).
% In this case, make sure you have the latest version of the \verb|hyperref| package (i.e.\ update your \LaTeX\ installation/packages) or, alternatively, ert back to \verb|\bibliographystyle{abbrv-doi}| (at the expense of removing hyperlinks from the bibliography) and try \verb|\bibliographystyle{abbrv-doi-hyperref}| again after some more editing.

\appendix
% \section*{Appendix A}
\begin{appendices}
\autoref{sec: metrics} presents computation details for the feature metrics used in \system, \autoref{sec: prompts} presents the prompts for the intelligent agents that provide feature recommendations and prompt suggestions, and \autoref{sec: experiment} presents an experiment to assess the consistency of LLMs as evaluators. All LLM-based implementations use OpenAI's ``gpt-3.5-turbo'' model.
The prompt block definitions and L2 definitions of each feature and used both in the interface and the prompts are listed in~\autoref{tab: block_definitions} and~\autoref{tab: feature_descriptions}.

\section{Feature Metrics}\label{sec: metrics}
\subsection{Complexity} 
% general desc then how people measure ling comp to measure readability, then intro score.
We define complexity as the opposite of readability, which is defined as the ease with which a reader can understand a written text. Following previous  works~\cite{novikova-etal-2017-need,kincaid1975derivation} in computational linguistics showing that readability is closely associated with the lexical diversity and structure of text,
we first calculate the readability of summaries using the Flesch Reading Ease Index~\cite{kincaid1975derivation}, a readability metric that measures the readability of a text based on sentence length and syllable count. The resulting score $r$ is a number between 0 and 100 that indicates the approximate educational level a person will need to be able to read a particular text easily, as shown in~\autoref{eq: Flesch}. Then we calculate the complexity score as $100-r$.  The Flesch Reading Ease Index provides an eight-level breakdown. For simplicity, we merge similar levels and generate a five-level breakdown, as given in~\autoref{tab: grade_Level}. 
Since we do not extend the formula, its limitation is also inherited: From \autoref{eq: Flesch}, we observe that $r$ relies on text length, and splitting long sentences into shorter ones can artificially decrease complexity. However, since LLMs don't inherently split sentences unless instructed, this limitation has minimal impact on our complexity calculation.

\begin{equation}
\label{eq: Flesch}
\begin{split}
    r = 206.835-1.015*(\frac{Total\_Words}{Total\_Sentences})-\\
    84.6*(\frac{Total\_Syllables}{Total\_Words})
\end{split}
\end{equation}

\begin{table}[h]
    \centering
    \begin{tabular}{|c|c|}
        \hline
        \textbf{Score} & \textbf{Level}\\ \hline
        90.0-100.0 & $Professional$\\ \hline  
        50.0-90.0 & $College$\\ \hline 
        40.0-50.0 & $High\ School$\\ \hline 
        10.0-40.0 & $Middle\ School$\\ \hline 
        0.0-10.0 & $Elementary$\\ \hline 
    \end{tabular}
    \caption{Categorization levels of complexity score}
    \label{tab: grade_Level}
\end{table}

\subsection{Formality}
Formality in language is defined as the degree of adherence to formal linguistic structures, conventions, and vocabulary. In natural language processing, a more formal language is often associated with a broader range of vocabulary~\cite{pu-demberg-2023-chatgpt}.
This is further supported by Heylighen et al.~\cite{heylighen1999formality} who associate formality with the explicitness of vocabulary.

We use the measure of
textual lexical diversity (MTLD)~\cite{mccarthy2010mtld} for formality calculation. 
MLTD measures the richness and diversity of vocabulary used in a text, considering both the frequency and the number of distinct words. It is calculated by averaging the length of consecutive word sequences in a text while ensuring they maintain a specific type-token ratio. It evaluates the text sequentially in both forward and backward passes to determine the average factor count, which represents the text's lexical variation. A higher MTLD score signifies a more formal text. 
% The authors do not provide a calculation formula for MTLD, but an implementation~\footnote{\url{https://github.com/kristopherkyle/lexical_diversity.git}} is available. 
We classify the final score as \textit{Informal, Standard, Formal, and Very Formal} based on the quantiles they lie in. Since we do not extend the formula, its limitations persist. MTLD's sensitivity to text length can cause inaccuracies in very short summaries due to significant partial factors. Additionally, its sequential processing may not fully capture the non-linear nature of human text comprehension.
\vspace*{-0.1cm}
\paragraph{\textit{\textbf{\footnotesize Computation Process}}}The text is processed sequentially from the beginning to the end.
A Type-Token Ratio (TTR) is calculated incrementally as each word is added to the sequence.
The TTR at each point is the ratio of the number of unique words (types) to the total number of words (tokens) up to that point. A factor is defined as a segment of the text where the TTR reaches a specified threshold, which is set at 0.72.
When the TTR reaches the threshold, one factor is counted, and the TTR calculation resets for the next segment. If the text ends before the TTR reaches the threshold for the last segment, a partial factor is calculated. Calculation of partial factors is given in~\autoref{eq: partial}. The total factor count is the sum of all complete and partial factors. Then the final MTLD score is given by~\autoref{eq: MTLD}. We directly adopted the implementation of the metric provided by the authors~\footnote{\url{https://github.com/kristopherkyle/lexical_diversity.git}}

\begin{equation}
\label{eq: partial}
\begin{split}
    Partial\_factor = \frac{1.00-TTR}{1.00-0.72}
\end{split}
\end{equation}

\begin{equation}
\label{eq: MTLD}
\begin{split}
    MTLD = \frac{\text{Total number of words}}{\text{Total Factor Count}}
\end{split}
\end{equation}

\subsection{Sentiment}
Sentiment refers to the emotional tone or attitude expressed in the text. 
We use VADER (Valence Aware Dictionary and sEntiment Reasoner)~\cite{hutto2014vader}, a lexicon and rule-based sentiment analysis tool that generates a sentiment score based on the words used, considering both the polarity (positive, negative, neutral) and strength of emotion. The calculation of VADER adopts a human-centric approach, where a valence-aware sentiment lexicon is constructed by human annotators. Also, it adopts five generalizable heuristics that reflect grammatical and syntactical patterns that humans use to express sentiment intensity. The final sentiment score is calculated by summing the valence scores of each word in the lexicon, adjusted according to the five heuristics. Since VADER calculates sentiment for each word individually, it may not accurately represent phrases, idioms, or other complex sentences and it's heavy reliance on lexicon quality and coverage may overlook misspellings, grammatical errors, and domain-specific words in overall sentiment calculation.  However, we use VADER over other deep learning (DL) models because DL models are often fine-tuned to a domain-specific dataset and lack transparency and robustness. A rule-based sentiment analyzer allows fast computation with consistent results. VADER also generalizes better, as it uses a number of well-established lexicon word banks and captures human perception of sentiment very well.
The sentiment score of a summary is given by~\autoref{eq: sentiment}
\begin{equation}
\label{eq: sentiment}
    \text{Sentiment Score} = \frac{\sum_{n=1}^{N} S_{i}}{N},
\end{equation}
where $S_{i}$ is the sentiment score of the $i^{th}$ word in the text, and N is the total number of words.

\subsection{Faithfulness}
Liu et al.~\cite{{liu2024cliqueparcel}} define faithfulness in natural language generation (NLG) as the degree to which the generated text is consistent with the input information or the reference text in terms of semantic similarity, completeness, and accuracy.
Although more advanced models exist, such as question generation and question answering (QG-QA) frameworks~\cite{durmus-etal-2020-feqa} and textual entailment models~\cite{maynez-etal-2020-faithfulness,falke-etal-2019-ranking}, we use \textbf{entity overlap} to measure faithfulness. It measures the overlap between the most important entities in the reference and generated texts. A high faithfulness score means the most important entities in the reference text are also present in the generated text. 
We use entity overlap instead of other state-of-the-art metrics because they tend to have slower computations and might introduce biases and randomness, with insignificant differences in performance. We prefer a robust metric devoid of randomness and offering fast computation speed for real-time analysis. 
\vspace*{-0.1cm}
\paragraph{\textit{\textbf{\footnotesize Computation Process}}}
The implementations of the whole process are outlined in Algorithm~\ref{alg: faith}.
We use the pre-trained SpaCy tagger~\cite{honnibal2017spacy} to extract entities from the article and the summary, and further incorporate an entity disambiguation step that groups similar entities with fuzzy string matching, as shown in Algorithm~\ref{alg: similar}. 
We first create a pair-wise similarity matrix between all entities.
Next, we organize these entities into disjoint sets, which are sets that do not share any common elements and have no intersection. 
Each disjoint set represents entities that have a fuzzy match ratio of at least $\epsilon$ within the set and less than $\epsilon$ when compared to entities in other disjoint sets. We set the value of $\epsilon$ as 0.7 in all our implementations. After grouping all the similar entities, we disambiguate the entities by only taking one unique entity from each disjoint set. Since not all entities in the article are important, Algorithm~\ref{alg: TopEnts} extracts important entities based on their frequency of occurrence using a fuzzy count. Although this is a naive ranking approach, we avoid using other supervised ranking algorithms for providing a fast and robust computation in real time. The most important entities extracted from the source are compared with the disambiguated entities in the summary. We then count the number of matches using the fuzzy match ratio. When there are very few (0--2) from the summary but many (more than 5) entities in the article, we adjust the score by comparing a fixed number of entities (Algorithm~\ref{alg: TopEnts}, Line \ref{penalty}). Finally, the score is calculated as the ratio of the number of matches to the number of most important entities in the article. We classify the final score as ``Bad'', ``Low'', ``Avg'' and ``Good'' based on the quantile they lie in. One significant limitation of our approach is its reliance on token-level entity overlap, which disregards the semantic context of the entities within the article and summary. While our method accommodates entity and synonym matching, it fails to consider paraphrasing and sentence-level similarity. To address this issue, we could employ deep learning-based sentence embedding models to evaluate the similarity between sentences in the article and the summary. However, in real-time applications, generating embeddings for lengthy texts can be both resource-intensive and time-consuming.
\begin{algorithm}[t]
\caption{Faithfulness score calculation}
\label{alg: faith}
\hspace*{\algorithmicindent} \textbf{Input}: article $a$, summary $s$,similarity threshold $\epsilon$ \\
\hspace*{\algorithmicindent} \textbf{Output}: Faithfulness score
\begin{algorithmic}[1]
\State $E_{s} \gets$ Entities in $s$
\State $E_{a} \gets$ Entities in $a$
% \State $L_{s} \gets len(E_{s})$
% \State $L_{a} \gets len(E_{a})$
% \State $S_{ij} = \begin{cases}
%     1 &  \text{fuzzy ratio}(E_{i},E_{j})\geq 0.7 \\
%     0 & \text{fuzzy ratio}(E_{i},E_{j})< 0.7
% \end{cases}$
\LeftComment{Similar entities in article and summary}
\State $S_{s} \gets$ SimilarEntites($E_{s},\epsilon$)
\State $S_{a} \gets$ SimilarEntites($E_{a},\epsilon$)
% \State $S_{s} \gets  O_{L_{s} \times L_{s}}$
% \State $S_{a} \gets  O_{L_{a} \times L_{a}}$
\State $D_{a} \gets $ Disjoint sets in $S_{s}$.
\State $D_{s} \gets $ Disjoint sets in $S_{a}$.
\LeftComment{Set of unique entities from each disjoint set} 
\State $U_{a} \gets$ Unique entities in $D_{a}$
\State $U_{s} \gets$ Unique entities in $D_{s}$
\State $T \gets$ GetTopEntites($U_{a},U_{s}$)
% \State $U_{a} \gets \{ (e_1, e_2,\ldots) \mid e_1 \in subset_{1}\subseteq D_{a}, e_2 \in subset_{2}\subseteq D_{a},\ldots \}$
% \State $U_{s} \gets \{ (e_1, e_2,\ldots) \mid e_1 \in subset_{1}\subseteq D_{s}, e_2 \in subset_{2}\subseteq D_{s},\ldots \}$
\LeftComment{Top source entity match count with unique summary entities.}
\State $M \gets \left| \{ (e_1, e_2) \mid e_1 \in U_s, e_2 \in T, \text{Fuzzy ratio}(e_1, e_2) \geq \epsilon \} \right|$

% \State $match\_count \gets$ MatchTopEntites($top\_article\_entities,U_{s}$)
\State $L_{T} \gets len(T)$
\State $score \gets \frac{M}{L_{T}}$ 

\end{algorithmic}
\end{algorithm}

\begin{algorithm}[t]
\caption{SimilarEntites($E,\epsilon$)}
\label{alg: similar}
\hspace*{\algorithmicindent} \textbf{Input:}List of all entities $E$, similarity threshold $\epsilon$ \\
\hspace*{\algorithmicindent} \textbf{Output:}Sets of similar entities $S$
\begin{algorithmic}[1]
\State $L_{e} \gets len(E)$
\State $S \gets O_{L_{e} \times L_{e}} $ \Comment{Zero matrix of size $L_{e} \times L_{e}$}
\For{$i \in rows(S)$ \& $j\in columns(S)$}
\State $S_{ij} = \begin{cases}
    1 &  \text{Fuzzy ratio}(E_{i},E_{j})\geq \epsilon \\
    0 & \text{Fuzzy ratio}(E_{i},E_{j})< \epsilon
\end{cases}$
\EndFor
\State \Return $S$
\end{algorithmic}
\end{algorithm}

\begin{algorithm}[t]
\caption{GetTopEntites($U_{a},U_{s}$)}
\label{alg: TopEnts}
\hspace*{\algorithmicindent} \textbf{Input:}Article entities $U_{a}$, summary entities $U_{s}$.\\
\hspace*{\algorithmicindent} \textbf{Output:}Most important entities in the article.
\begin{algorithmic}[1]

\State Sort$(U_{a})$, descending by their fuzzy occurrence in article.
\State $L_{s} \gets len(U_{s})$
\State $L_{a} \gets len(U_{a})$
\If{$L_{s}<3$ and $L_{a}>4$} \label{penalty}
    \State $top\_article\_entities \gets U_{a}[0:5]$          
\ElsIf{$L_{a}<L_{s}$ or $L_{s}=0$}
    \State $top\_article\_entities \gets U_{a}$        
\Else 
\State $top\_article\_entities\gets U_{a}[0:L_{s}]$
\EndIf
\State \Return $top\_article\_entities$
\end{algorithmic}
\end{algorithm}

\subsection{Naturalness}
The naturalness of text is a measure of how well the text flows, sounds like something a native speaker would produce, is easy to understand, and adheres to the rules of the language. 
To the best of our knowledge, no known metrics were proposed to measure the naturalness of a text. In this work, we measure naturalness based on linguistic feature differences observed from human and LLM-generated text.
All subsequent experiments are conducted on the dataset released by Zhang et al.~\cite{zhang2024benchmarking}. It is important to note that our approach, tested on general-domain articles, may not generalize to other tasks, domains, or languages. Its applicability in scientific terminology and languages with different grammatical and semantic structures remains unexplored, where other linguistic features might be more relevant. Additionally, this study does not cover NLG tasks like question-answering and logical reasoning. However, similar methods could be applied to these tasks, domains, and languages, as analyzing a set of features has shown better alignment with human judgments in various studies~\cite{groves2018treat, munoz2023contrasting, temperley2008dependency}.

\subsubsection{Differentiating Linguistic Features}\label{subsec: exp}
% We extract linguistic features from human and LLM summaries with a dataset developed to benchmark LLM-generated summaries~\cite{zhang2024benchmarking}. 
% We assess the importance of each linguistic feature using PCA and assign trained weights using a random forest classifier. The weighted average of these features is used to obtain the final naturalness score.
 Inspired by the findings of previous works~\cite{groves2018treat, munoz2023contrasting} that human and LLM-generated summaries exhibit different statistical patterns on certain linguist features, we measure the naturalness of text by those that can differentiate human and LLM-generated texts. As LLMs advance quickly, we sought to verify their findings on state-of-the-art LLMs and selected a set of linguistic features to experiment with, namely average arc lengths, average subtree heights, average dependency-tree heights, average sentence lengths, and average word lengths. First, we computed these features on all the summaries in the dataset and conducted PCA on these features, as shown in~\autoref{tab: pca}.
\begin{table}[htbp]
  \centering
  \caption{Importance assigned by PCA}
    \begin{tabular}{lr}
    \toprule
    \textbf{Feature} & \textbf{Importance} \\
    \midrule
     \cellcolor{gray!25}avg\_dependency\_tree\_heights & \cellcolor{gray!25}0.504439 \\
     \cellcolor{gray!25}avg\_left\_subtree\_height & \cellcolor{gray!25}0.486318 \\
     \cellcolor{gray!25}avg\_right\_subtree\_height & \cellcolor{gray!25}0.441374 \\
     \cellcolor{gray!25}average\_sentence\_lengths & \cellcolor{gray!25}0.427449 \\
    num\_right\_subtrees & 0.252828 \\
    num\_left\_subtrees & 0.158020 \\
    avg\_total\_arc\_length & 0.137703 \\
    avg\_right\_arc\_length & 0.128035 \\
    avg\_left\_arc\_length & -0.081986 \\
    avg\_word\_lengths & -0.023296 \\
    \bottomrule
    \end{tabular}%
  \label{tab: pca}%
\end{table}%

To better understand the PCA results, we generated visualizations to explain the differentiating capability of each feature. We observed that the number of subtrees (right and left), average arc lengths (total, right, and left), and average word lengths do not show any significant difference.  Next, we present the observed differences in other features.

\begin{figure*}[!t]
    \begin{minipage}{\textwidth}
        \centering
        \includegraphics[width=0.49\linewidth]{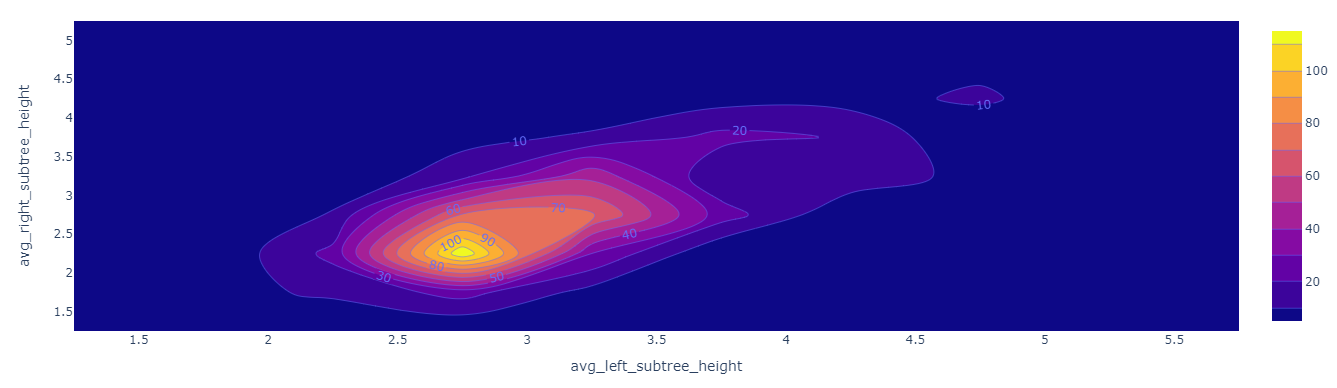} 
        \hfill 
        \includegraphics[width=0.49\linewidth]{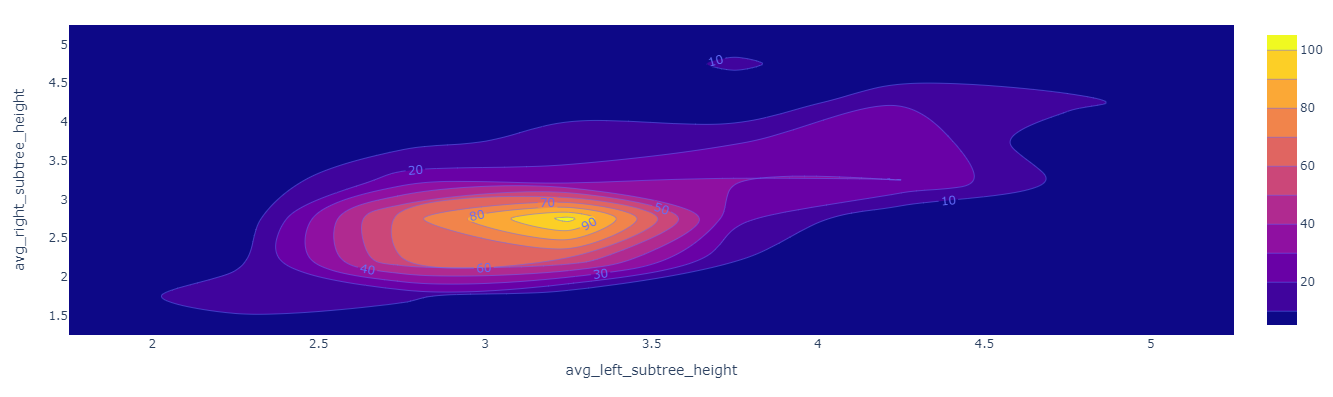} 
        \caption{
        This contour plot compares the average left vs right subtree heights of human (left) and llm-generated (right) summaries. The color signifies the number of summaries with the corresponding subtree lengths, with lighter shades demonstrating more number of summaries.
        Human written summaries have shorter average left, and right subtree heights as compared to LLM written summaries. 
        This conclusion comes from the observation that the densest point (yellow region) for the LLM summaries is closer to the lower left origin than human summaries.
        It is also observed that human-generated summaries consistently exhibit a higher frequency of balanced left and right subtrees compared to LLM summaries. Specifically, there are 182 balanced human summaries and 133 balanced LLM summaries.}
        \label{fig: subtree}
    \end{minipage}
\end{figure*}

\vspace*{-0.1cm}
\paragraph{\textit{\textbf{\footnotesize Subtree Heights}}}
An analysis of the subtrees in the dependency parse trees of summaries shown in~\autoref{fig: subtree} revealed that humans tend to produce text with balanced left and right subtrees more frequently as compared to LLM summaries. This is also coherent with the works of Temperley et al.~\cite{temperley2008dependency} which shows that sentences that have a balanced depth on either side of the root in the dependency parse tree are judged to be more natural and well-worded by humans. We also note that human-written summaries tend to have shorter average left and right sub-tree heights more frequently than LLM summaries. These observations indicate that texts with shorter and balanced sub-tree heights are more likely to exhibit human-like naturalness.

\begin{figure*}[!t]
    \begin{minipage}{\textwidth}
        \centering
        {\includegraphics[width=0.49\textwidth]{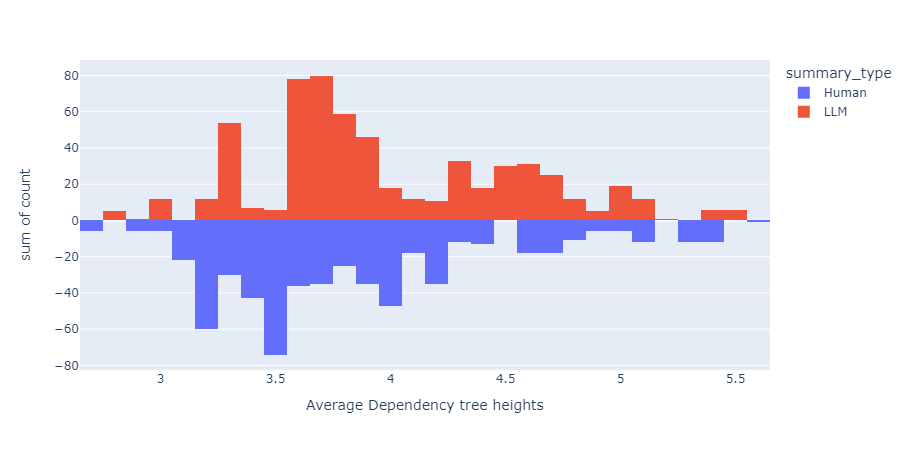}}
        \hfill 
        {\includegraphics[width=0.49\textwidth]{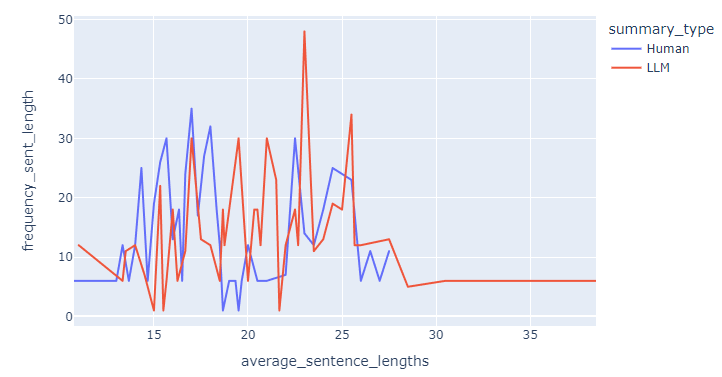}}
        \caption{\textbf{(a)}  Average dependency tree heights vs sum of counts of summaries with the respective height. LLM tends to produce very few summaries with short dependency tree heights with only 97 summaries out of 599 having a dependency tree height of 3.5 or lower as compared to 247 human summaries. Human summaries have less frequency of average dependency tree heights beyond 3.5. This highlights that human text exhibits simpler and shorter dependencies more often.
        \textbf{(b)}  Average sentence lengths vs frequency of summaries with the respective length. The blue line denotes humans and the red denotes LLM summaries. It is observed that LLM summaries tend to have longer average sentence lengths beyond the length of 18. This indicates that human summaries have fewer instances with longer average sentence lengths as compared to LLM summaries.}
        \label{fig: abg_dep_tree_and_sent_len}
    \end{minipage}
\end{figure*}

\vspace*{-0.1cm}
\paragraph{\textit{\textbf{\footnotesize Dependency Tree Heights }}}
We observe that humans tend to produce summaries of shallower dependency tree heights more often than LLMs.
In natural language, deeper dependency parse trees have more linguistic and dependency complexity. Groves et al.~\cite{groves2018treat} train classifiers on word embeddings and linguistic features to weigh their importance in the automatic evaluation of naturalness. They find dependency tree height to be one of the most important linguistic features. Following this insight, we compare the dependency tree heights of LLM and human summaries in~\autoref{fig: abg_dep_tree_and_sent_len}-a. LLM tends to produce very few summaries
with short (<3.5) dependency tree heights, while human summaries have less frequency of long (>3.5)
dependency tree heights. 
This indicates that text with shorter dependencies and lesser linguistic complexity correlates better with humans.
\vspace*{-0.1cm}
\paragraph{\textit{\textbf{\footnotesize Sentence Length}}}
We further compare the average sentence lengths of human and LLM summaries in~\autoref{fig: abg_dep_tree_and_sent_len}-b which shows that humans tend to produce lesser summaries with longer average sentence lengths as compared to LLM summaries. Human texts also have shorter sentence lengths more frequently. Conversely, LLM tends to generate longer sentences more frequently, which can be attributed to their tendency to hallucinate in some cases.

\subsubsection{Naturalness Score Calculation}
Based on previous works~\cite{temperley2008dependency,groves2018treat} and the aforementioned experiments, we identified a set of linguistic and lexical features that have been shown to exhibit statistical differences between human and LLM-generated texts, consisting of average subtree height (right and left), average dependency tree height, and average sentence length. This feature set is used to train a random forest classifier. The classification task is to predict if a summary is written by a human or an LLM based on the linguistic features. The classifier achieves an accuracy of $99.6\%$ on a test size of 300 human or LLM-generated texts~\cite{zhang2024benchmarking}. The final weights assigned to these metrics by the classifier are used to calculate the weighted average of the feature values in~\autoref{eq: UnscaledScore}:
\begin{equation}
\label{eq: UnscaledScore}
    \begin{split}
    &\bar{X}=\frac{\sum_{n=1}^{N} W_{i}*x_{i}}{N}
\end{split}
\end{equation}
$W_{i}$ is the weight of the $i^{th}$ feature, $x_{i}$ is the feature value and $N$ is the number of features.
In~\autoref{eq: natural}, we normalize the score and invert it to better align with human judgment (higher means more natural). We classify the final score as ``Bad'', ``Low'', ``Avg'' and ``Good'' based on the quantiles they lie in.
\begin{equation}
\label{eq: natural}
    \begin{split}
    &naturalness\_score=1-\frac{\bar{X}-min(\bar{X})}{max(\bar{X})-min(\bar{X})}
\end{split}
\end{equation}

\clearpage
\section{Prompts}\label{sec: prompts}
The intelligent agents for supporting prompt suggestions and feature recommendation are implemented with a template prompt using OpenAI's ``gpt-3.5-turbo'' model. Each template consists of two parts: a system prompt that describes the task and provides necessary contexts, such as feature definitions and prompt block definitions, and a user prompt, which is injected with the user's input. Below, we provide the templates for prompt suggestions and feature recommendations.
\subsection{Prompt Suggestions}
The prompt for prompt suggestions takes three inputs: \textit{block\_name}, \textit{block\_definition}, 
as listed in~\autoref{tab: block_definitions} and user question (\textit{question}):

\begin{figure}[h!]
\centering
\includegraphics[width=\columnwidth]{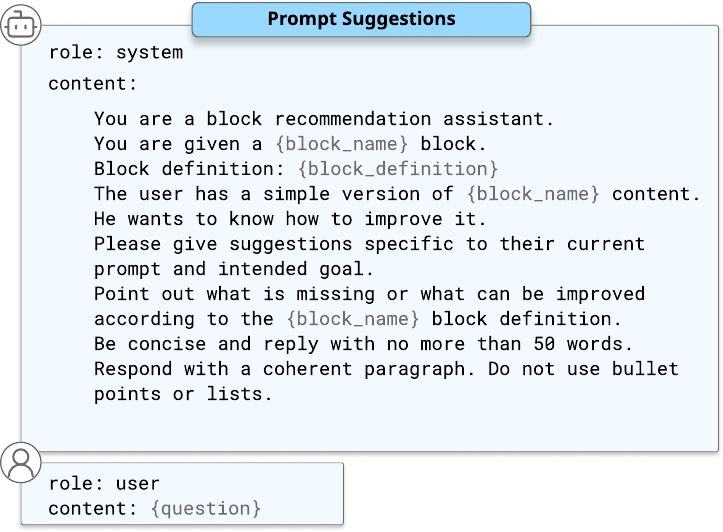} % 
        \caption{
        Prompt template for providing prompt suggestions. Grayed-out words are variables to be injected.}
        \label{fig: p_prompt_suggestions}
\end{figure}

\begin{table}[h!]
    \centering
    \includegraphics[width=\textwidth]{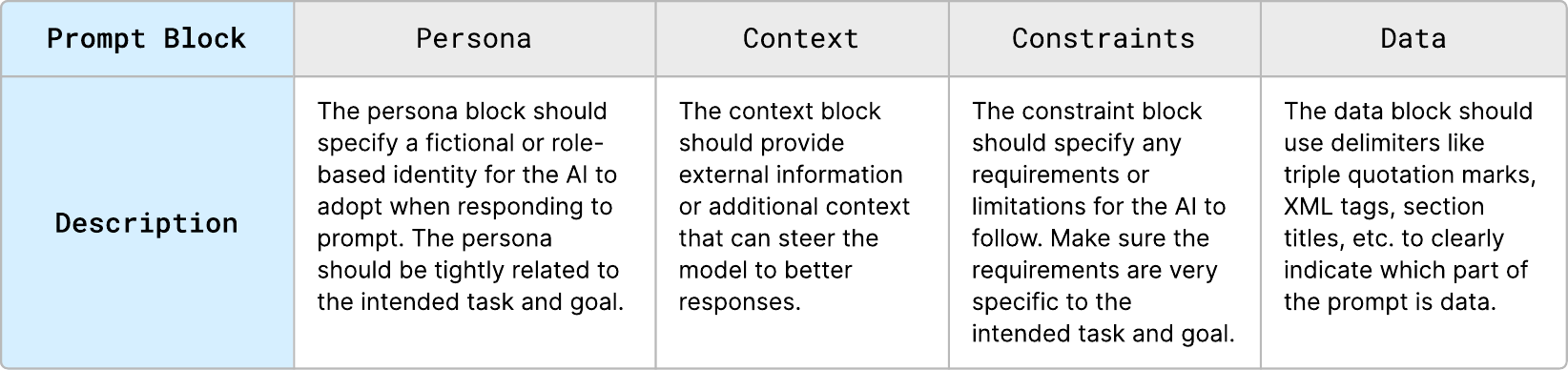}
   \captionsetup{width=\textwidth,justification=centering}
\caption{Prompt block definitions derived from prompting methodologies.}
\label{tab: block_definitions}
\end{table}

\newpage

\begin{table*}
\begin{center}
\scalebox{1.1}{%
\begin{tabular}{|c|p{5cm}|c|c|p{4cm}|}
\hline
\rowcolor{LightCyan}
\mc{1}{\textbf{Metric}} & \mc{1}{\textbf{Description}} & \mc{1}{\textbf{Range}} & \mc{1}{\textbf{Level}} & \mc{1}{\textbf{Notes}}\\ \hline

\multirow{12}{*}{Complexity} &  \multirow{12}{*}{\parbox{5cm}{Complexity metrics aim to quantify the readability of a piece of writing by considering various linguistic features, such as sentence length, word length, syllable count or semantic difficulty}} &  \multirow{2}{*}{0, 10} & \multirow{2}{*}{Elementary} & Very easy to read. Easily understood by an 11-year-old student  \\ \cline{3-5}
& & \multirow{3}{*}{10, 40} & \multirow{3}{*}{Middle School} & Plain English. Easily understood by 13- to 15-year-old students.\\ \cline{3-5}
& & \multirow{2}{*}{40, 50} & \multirow{2}{*}{High School} & Fairly difficult to read. Best understood by high school students\\ \cline{3-5}
& & \multirow{2}{*}{50, 90} & \multirow{2}{*}{College} & Very difficult to read. Best understood by college students.\\ \cline{3-5}
& & \multirow{3}{*}{90, 100} & \multirow{3}{*}{Professional} & Extremely difficult to read. Best understood by university graduates.\\ \hline

\multirow{5}{*}{Formality} &  \multirow{4}{*}{\parbox{5cm}{Formality measures how formal a piece of writing is.}} &  
\multirow{2}{*}{0, 60} & \multirow{2}{*}{Informal} & Casual and conversational language.\\ \cline{3-5}
& & \rule{0pt}{2.5ex}60, 100 & \rule{0pt}{2.5ex}Standard & \rule{0pt}{2.5ex}Standard, neutral language.\\ \cline{3-5}
& & \multirow{2}{*}{100, 200} & \multirow{2}{*}{Formal} & Professional or academic language\\ \cline{3-5}
& & \multirow{2}{*}{200, -1} & \multirow{2}{*}{Very Formal} & Highly technical or legal language\\ \hline

\multirow{3}{*}{Sentiment} &  \multirow{3}{*}{\parbox{5cm}{Sentiment aims to determine the attitude of a writer with respect to the overall contextual polarity.}} & \rule{0pt}{2.5ex} -1, -0.3 & \rule{0pt}{2.5ex}Negative & \rule{0pt}{2.5ex}Negative sentiment\\ \cline{3-5}
& & \rule{0pt}{2.5ex}-0.3, 0.3 & \rule{0pt}{2.5ex}Neutral & \rule{0pt}{2.5ex}Neutral sentiment\\ \cline{3-5}
& & \rule{0pt}{2.5ex}0.3, 1 & \rule{0pt}{2.5ex}Positive & \rule{0pt}{2.5ex}Positive sentiment\\
\hline

\multirow{14}{*}{Faithfulness} &  \multirow{14}{*}{\parbox{5cm}{Faithfulness measures how broad and accurate are the generated summary using the name entity overlap between the summary and the article. Common named entities include persons, organizations, locations, or dates.}} &  \multirow{4}{*}{0.0, 0.25} & \multirow{4}{*}{Bad}& All or Almost all of the important entities from the article are missing in the generated summary\\ \cline{3-5}
& & \multirow{3}{*}{0.25, 0.4} & \multirow{3}{*}{Low} & Very few of the important entities from the article are present in the generated summary\\ \cline{3-5}
& & \multirow{3}{*}{0.4, 0.6} & \multirow{3}{*}{Avg} & Few of the important entities from the article are missing in the generated summary\\ \cline{3-5}
& & \multirow{4}{*}{0.6, 1.0} & \multirow{4}{*}{Good} & All or almost all the important entities from the article are present in the generated summary\\ 
\hline

\multirow{8}{*}{Naturalness} &  \multirow{8}{*}{\parbox{5cm}{Naturalness of text is a measure of how well the text flows and sounds like something a native speaker would produce. It should be easy to understand and should adhere to the rules of the language}} &  \multirow{2}{*}{0.0, 0.435} & \multirow{2}{*}{Bad}& Summary is not natural or human-like\\ \cline{3-5}
& & \multirow{2}{*}{0.435, 0.607} & \multirow{2}{*}{Low} & Summary is somewhat natural and human-like\\ \cline{3-5}
& & \multirow{2}{*}{0.607, 0.715} & \multirow{2}{*}{Avg} & Summary is mostly natural and human-like\\ \cline{3-5}
& & \multirow{2}{*}{0.715, 1.0} & \multirow{2}{*}{Good} & Summary is natural and human-like\\ 
\hline

\multirow{4}{*}{Length} &  \multirow{4}{*}{\parbox{5cm}{Length measures the number of words in the summary.}} &  \rule{0pt}{2.5ex}0, 100 & \rule{0pt}{2.5ex}Short& \rule{0pt}{2.5ex}Under 100 words.\\ \cline{3-5}
& & \rule{0pt}{2.5ex}100, 300 & \rule{0pt}{2.5ex}Mid & \rule{0pt}{2.5ex}100 to 300 words.\\ \cline{3-5}
& & \rule{0pt}{2.5ex}300, 500 & \rule{0pt}{2.5ex}Long & \rule{0pt}{2.5ex}300 to 500 words.\\ \cline{3-5}
& & \rule{0pt}{2.5ex}500, -1 & \rule{0pt}{2.5ex}Very long & \rule{0pt}{2.5ex}Over 500 words.\\ 
\hline
\end{tabular}}
\end{center}
\caption{Feature descriptions. In the prompt, we combine \textit{Metrics}, \textit{Description}, \textit{Levels} and \textit{Notes} with a string template.}
\label{tab: feature_descriptions}
\end{table*}

\subsection{Feature Recommendation}
The prompt for feature recommendation takes two inputs: \textit{feature\_descriptions} as listed in~\autoref{tab: feature_descriptions} and user question (\textit{question}), and uses the JSON mode functionality to define a structured output:
\begin{figure}[h!]
\centering
\includegraphics[width=\columnwidth]{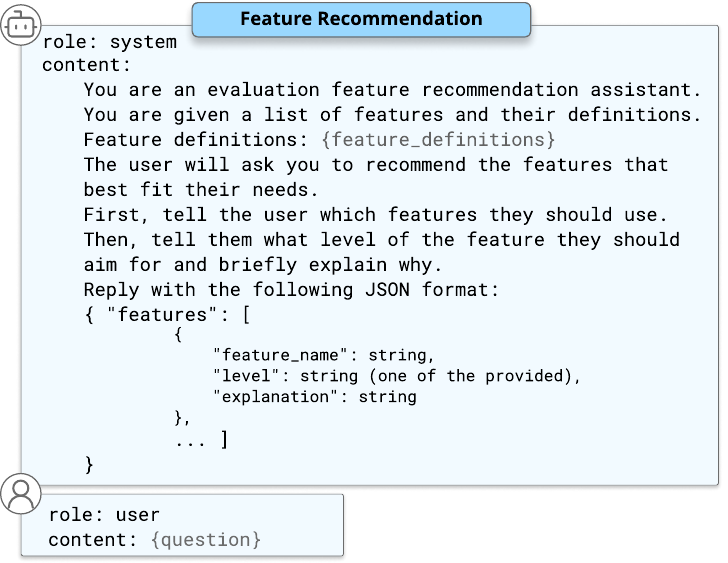} % 
        \caption{
        Prompt template for providing feature recommendations. Grayed-out words are variables to be injected.}
        \label{fig: p_feature_recommendation}
\end{figure}

\newpage

\begin{figure*}
    \centering
    \begin{minipage}{0.49\textwidth}
        \centering
        \includegraphics[width=\linewidth]{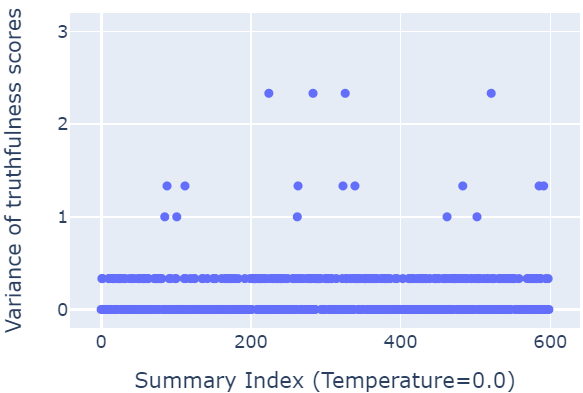}
    \end{minipage}
    \hfill
    \begin{minipage}{0.49\textwidth}
        \centering
        \includegraphics[width=\linewidth]{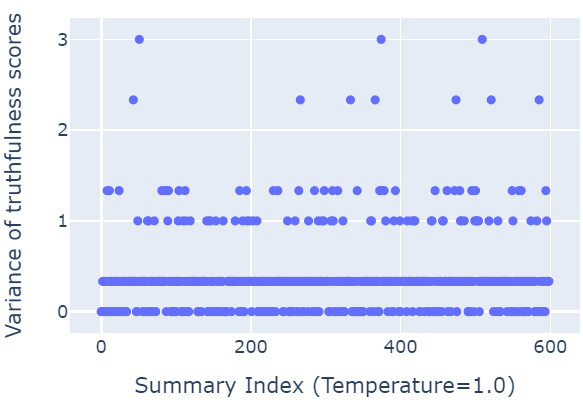}
    \end{minipage}
    
    \caption{x-axis: individual summaries (indexed from 0 to 599), y-axis: variance in truthfulness scores. For each summary, we used three levels of definitions (no definition, beginner, and expert) to generate a truthfulness score, and calculated the variance (y-axis). We varied the temperature setting (0, 0.3, 0.7, 1.0) to ensure that the variance is persistent. At temperature 0.0 (left), we see that the majority of summaries have variance below 0.5. At temperature 1.0 (right), the variances are enlarged. The experiment shows that even at its most consistent setting (temperature=0), the LLM still shows variance in the generated scores when the metric definitions have varying levels of detail.}
    \label{fig:truthfulness}
\end{figure*}

\begin{figure*}
    \centering
    \begin{minipage}{0.49\textwidth}
        \centering
        \includegraphics[width=\linewidth]{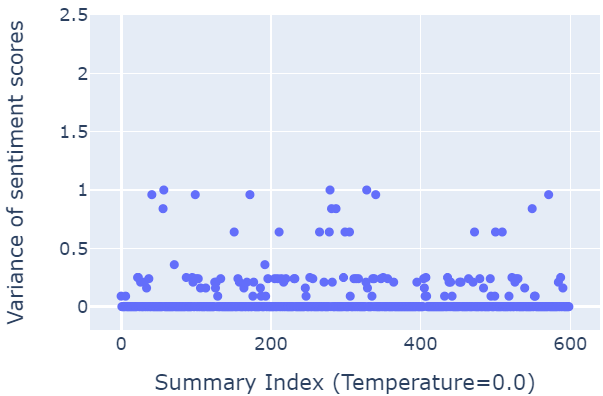}
    \end{minipage}
    \hfill
    \begin{minipage}{0.49\textwidth}
        \centering
        \includegraphics[width=\linewidth]{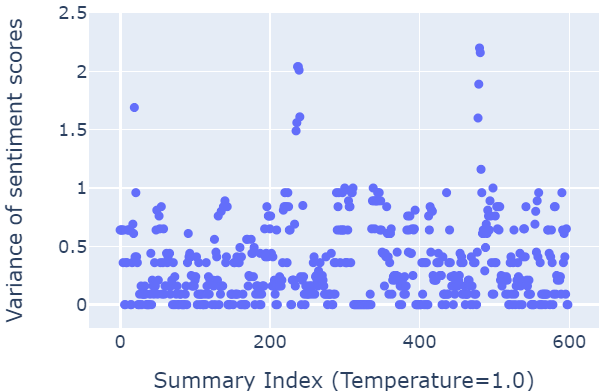}
    \end{minipage}
    
    \caption{x-axis: individual summary index (indexed from 0 to 599), y-axis: variance in sentiment scores. For each summary, we ran the same prompt ten times and calculated the variance in the ten generated scores. We varied the temperature setting (0, 0.5, 1.0) to ensure that the variance is persistent. 
    We observe that even at temperature 0.0 (left), a significant number of summaries have inconsistent scores, and the most inconsistent ones have variance=1.0.
    At temperature 1.0 (right), the variances are enlarged.
    }
    \label{fig:iterations}
\end{figure*}

\section{Experiment: LLMs as evaluators}\label{sec: experiment}
Recently, LLM evaluators~\cite{wang2023pandalm, zheng2023llmjudge, kim2023evallm}, where LLMs are prompted to output a numerical evaluation score, have become more popular due to their low technical barrier and high customizability in defining an evaluation criterion. 
However, previous works~\cite{lu-etal-2022-fantastically, ouyang2023llm} have shown that LLMs are non-deterministic and sensitive to slight changes in prompts in diverse NLP contexts, such as sentiment classification, code generation, and text summarization. Shen et al.~\cite{shen2023large} concluded that LLMs are not robust enough as human level evaluators. 
Building upon these works, our experiment is designed to explore, from the end-user's perspective, how consistent LLMs are as evaluator in text summarization. We seek to answer the following research questions:
\begin{itemize}
\item \textbf{RQ1:} How do minor variations in the prompt defining an evaluation metric affect the variability of the metric score?
\item \textbf{RQ2:} Do LLMs maintain consistency in metric scores when the same prompt defining an evaluation metric is applied repeatedly?
\end{itemize}

\paragraph{\textit{\textbf{\footnotesize Procedure}}} 
To investigate \textbf{RQ1}, we chose three evaluation metrics: sentiment, readability, and truthfulness, ranging from 1 to 5. For each metric, we created three prompts with varying levels of definition: (1) no definition, (2) a beginner-level, non-technical definition, and (3) an expert-level, detailed definition. A prompt with no definition of evaluation criterion represents prompt designers with no prior knowledge of natural language evaluation. The beginner-level prompt provides a naive way of defining an evaluation metric representative of non-technical prompt designers. The expert-level prompt gives a detailed definition of the metric with breakdown of scores to better align the responses according to the user's intent, representing experienced prompt designers. 
These prompts were then used to evaluate all summaries in the dataset. We calculated the variance between the three definitions to inspect the sensitivity of LLMs.
The same procedure is repeated under four temperature settings: 0.0, 0.3, 0.7, and 1.0.

To investigate \textbf{RQ2}, We chose two evaluation metrics, sentiment and readability, each with a single prompt that includes a beginner-level definition of the metric, outputing a score ranging from 1 to 5. For each metric, we repeatedly applied the same prompt to the dataset 10 times, and then examined the variance in the generated scores.
The same procedure is repeated under three temperature settings of 0.0, 0.5, and 1.0.
The specific prompts employed in these experiments are documented in~\autoref{tab:prompts}.

Our experiment utilizes the dataset released by Zhang et al.~\cite{zhang2024benchmarking} comprising 599 articles and corresponding summaries generated by LLMs. For sentiment and readability, we only assess the summaries. For truthfulness, we take each pair of article and summary as input. 
For the model, We utilized OpenAI’s ``gpt-3.5-turbo'' for generation of all metric scores. We do not use GPT-4 due to its cost of usage and rate limit restrictions. Moreover, the non-determinism of GPT-4 as compared to GPT-3.5-turbo remains to be persistent~\cite{ouyang2023llm}.

\paragraph{\textit{\textbf{\footnotesize Conclusion and discussion}}}  
We analyze the variance in metric scores across the three levels of definitions to address \textbf{RQ1}. From~\autoref{fig:truthfulness}, we observe the inconsistency in the truthfulness scores for three different prompts defining a single metric. 
The temperature parameter balances the consistency-creativity trade-off, with 0.0 temperature being their most consistent setting and 1.0 the most creative.  
However, even at 0.0 temperature, the variance across the three truthfulness scores assigned by the three definitions is non-zero for a significant amount of summaries, highlighting the inconsistencies in scoring. The variance increases as temperature is increased, with 1.0 temperature setting having the highest variance. Similar patterns are observed for readability and sentiment. This indicates that variations in the evaluation prompts can lead to variance in scores. Consequently, allowing users to define new evaluation metrics can introduce more uncertainty to the system. The exact variance in scores for the three metrics across the three definitions for the whole dataset is documented in~\autoref{tab:variance}. 

To address \textbf{RQ2}, we analyze the variance of scores when the same prompt is applied multiple times to the same dataset. From~\autoref{fig:iterations}, we observe that even at 0.0 temperature, the variance of sentiment scores is non-zero for a significant number of summaries. As the temperature is increased, we observe a higher variance in scores and more number of summaries with non-zero variance, with 1.0 temperature setting being the most inconsistent. We observe similar patterns in readability. This indicates that the application of an identical prompt to the same group of summaries can yield varying scores, leading to concerns about the reliability of the results amongst users. The exact variances observed are given in~\autoref{tab:var_iterations}.

\begin{table}[h!]
    \centering
    \begin{tabular}{|c|c|c|c|c|}
        \hline
        \multirow{2}{*}{Metric} & \multicolumn{4}{c|}{Temperature} \\ \cline{2-5}
         & 0.0 & 0.3 & 0.7 & 1.0  \\ \hline
        
        Sentiment  & 0.192 & 0.197 & 0.196 & 0.222 \\ \hline
        Readability & 0.104 & 0.103 & 0.131 & 0.160  \\ \hline
        Truthfulness & 0.110 & 0.138 & 0.207 & 0.265  \\
        \hline
    \end{tabular}
    \caption{Variance of Sentiment, Readability, and Truthfulness measured across the scores generated by the three levels of definitions at different temperatures.}
    \label{tab:variance}
\end{table}

\begin{table}[h!]
    \centering
    \begin{tabular}{|c|c|c|c|}
        \hline
        \multirow{2}{*}{Metric} & \multicolumn{3}{c|}{Temperature} \\ \cline{2-4}
         & 0.0 & 0.5 & 1.0  \\ \hline
        
        Sentiment  & 0.057 & 0.174 & 0.334  \\ \hline
        Readability & 0.025 & 0.124 & 0.191   \\ \hline
    \end{tabular}
    \caption{Variance of scores for Sentiment and Readability measured across the ten iterations of the same dataset.}
    \label{tab:var_iterations}
\end{table}

\paragraph{\textit{\textbf{\footnotesize Limitations}}} Given the resource constraints, our experiment was conducted at a rather small scale, using a single dataset, 10 iterations, three different metrics, and four temperature settings. For a more robust conclusion, the experiment can be extended to cover more diverse datasets and hyperparameter settings. 
Also, at a larger scale, a comprehensive statistical analysis can be added to examine the robustness of the experiment. For this experiment, we targeted the application scenario in the \system 
system to more robustly show the inapplicability of LLM evaluators, and leave a more comprehensive evaluation of LLM evaluators for future research.

\clearpage
\begin{table*}[ht]
    \centering
    \small
    \begin{tabular}{|c|c|p{\textwidth-4cm}|}
        \hline
        \rowcolor{LightCyan}
        \mc{1}{\textbf{Metric}} & \mc{1}{\textbf{Prompt Definition}} &\mc{1}{\textbf{Prompt }}\\ \hline
        \multirow{21}{*}{Sentiment}& \multirow{5}{*}{No Definition} & role: system \newline content: The user will provide a summary to be evaluated. 
                Evaluate the sentiment of the summary. 
                Reply strictly with a single digit score from this list of scores: 1: Negative, 2: Slightly Negative, 3: Neutral, 4: Slightly Positive, 5: Positive.
                \newline role: user \newline content: \textcolor{gray}{\{Summary\}}\\
        \cline{2-3}
        & \multirow{5}{*}{Beginner} & role: system \newline content: The user will provide a summary to be evaluated. 
        The metric to be evaluated is sentiment, which refers to the emotional tone or attitude expressed within a given text.
                Evaluate the sentiment of the summary. 
                Reply strictly with a single digit score from this list of scores: 1: Negative, 2: Slightly Negative, 3: Neutral, 4: Slightly Positive, 5: Positive.
                \ role: user \newline content: \textcolor{gray}{\{Summary\}}\\
        \cline{2-3}
        &\multirow{11}{*}{Expert} & role: system \newline content: The user will provide a summary to be evaluated. 
                Sentiment refers to the emotional tone or attitude expressed within a given text.
                Evaluate the sentiment of the summary. 
                The possible scores of sentiment are:\newline
                1: Negative, Strongly unfavorable or pessimistic tone. Expresses disapproval, dissatisfaction, or criticism.\newline
                2: Mildly negative sentiment. Indicates some negativity, but not as intense as the “Negative” category.\newline
                3: Neutral, Neither positive nor negative. Presents information objectively without emotional bias.\newline
                4: Mildly positive sentiment. Conveys a favorable or optimistic tone, but not strongly so.\newline
                5: Positive, Strongly favorable or enthusiastic. Expresses approval, satisfaction, or delight.\newline
                Reply strictly with a single digit score.
                \newline role: user \newline content: \textcolor{gray}{\{Summary\}}\\
        \hline
        \multirow{23}{*}{Readability}&\multirow{5}{*}{No definition} & role: system \newline content: The user will provide a summary to be evaluated. 
                Evaluate the readability of the summary. 
                You need to assign a score on a scale of 1 to 5.
                Reply strictly with a single digit score.
                \newline role: user \newline content: \textcolor{gray}{\{Summary\}}\\
        \cline{2-3}
        &\multirow{6}{*}{Beginner} & role: system \newline content: The user will provide a summary to be evaluated. 
                The metric to be evaluated is readability, which is defined as the ease with which a human can understand a written text.
                Evaluate the readability of the summary. 
                You need to assign a score on a scale of 1 to 5.
                Reply strictly with a single digit score.  
                \newline role: user \newline content: \textcolor{gray}{\{Summary\}}\\
        \cline{2-3}
        &\multirow{11}{*}{Expert} & role: system \newline content: The user will provide a summary to be evaluated. 
                Readability is defined as the ease with which a human can understand a written text.
                Evaluate the readability of the summary. 
                You need to assign a score on a scale of 1 to 5.
                The breakdown of the scores is as follows:\newline
                1: Very difficult to read, but can be read by professionals.\newline
                2: Difficult to read, but can be read by well-read individuals.\newline
                3: Fairly difficult to read but can be read by middle-aged students\newline
                4: Easy to read, can be read by most people\newline
                5: Very easy to read\newline
                Reply strictly with a single digit score.  
                \newline role: user \newline content: \textcolor{gray}{\{Summary\}}\\
        \hline
        \multirow{25}{*}{Truthfulness}&\multirow{5}{*}{No definition} & role: system \newline content: The user will provide an article and a summary for that article. 
                Evaluate the truthfulness of the summary, with the article as the ground truth. You need to assign a score on a scale of 1 to 5.
                Reply strictly with a single digit score.
                \newline role: user \newline content: \textcolor{gray}{\{Article, Summary\}}\\
        \cline{2-3}
        &\multirow{6}{*}{Beginner} & role: system \newline content: The user will provide an article and a summary for that article. 
                Evaluate the truthfulness of the summary with the article as the ground truth.
                Truthfullness refers to how accurately the summary reflects the actual facts and content of the original article. 
                You need to assign a score on a scale of 1 to 5.
                Reply strictly with a single digit score. 
                \newline role: user \newline content: \textcolor{gray}{\{Article, Summary\}}\\
        \cline{2-3}
        &\multirow{11}{*}{Expert} & role: system \newline content: The user will provide an article and a summary for that article. 
                Evaluate the truthfulness of the summary, with the article as the ground truth. You need to assign a score on a scale of 1 to 5.
                The breakdown of scores is as follows:\newline
                1: Not Truthful: Information given is completely false or misleading, no element of truth.\newline
                2: Slightly Truthful: Information has small elements of truth, but is largely flawed or inaccurate, significant omissions of relevant facts.\newline
                3: Moderately Truthful: Information is a mix of truth and inaccuracies. Its truthfulness may depend on the interpretation or context.\newline
                4: Mostly Truthful: high level of truth. May contain minor errors or omissions, or slightly misleading.\newline
                5: Completely Truthful: The information presented is entirely accurate, factual, and free of any misleading elements or omissions.\newline
                Reply strictly with a single digit score. 
                \newline role: user \newline content: \textcolor{gray}{\{Article, Summary\}}\\
        \hline
    \end{tabular}
    \caption{Prompts with different levels of definitions for different feature metrics, each prompt used as a separate evaluation system for observing variance in scores assigned by LLMs. Grayed-out words are inputs to the prompt.}
    \label{tab:prompts}
\end{table*}

\end{appendices}

\end{document}